\documentclass[hidelinks,12pt]{article}
\usepackage{graphicx}
\usepackage{amssymb}
\usepackage{amsmath}
\usepackage[hypertexnames=false]{hyperref}
\hypersetup{
    colorlinks=true,
	linkcolor=blue,
	urlcolor=blue,
    citecolor=blue
}
\usepackage{braket, cancel}
\usepackage[capitalise,noabbrev,nameinlink]{cleveref}
\numberwithin{equation}{section}
\pdfoutput=1

\def\gsim{\, \rlap{$>$}{\lower 1.1ex\hbox{$\sim$}}\,}
\def\lsim{\, \rlap{$<$}{\lower 1.1ex\hbox{$\sim$}}\,}
\newcommand{\comment}[1]{}

\newcommand{\mbf}[1]{\mathbf #1}

\def\x{\mbf x}

\newcommand{\be}{\begin{equation}}
\newcommand{\ee}{\end{equation}}
\newcommand{\bea}{\begin{eqnarray}}
\newcommand{\eea}{\end{eqnarray}}

\begin{document}

\begin{titlepage}

\setcounter{page}{1} \baselineskip=15.5pt \thispagestyle{empty}

\vspace{0.5cm}
\begin{center}
{\Large \sc
Cosmological Particle Production\\
without Quantum Fields\\
\vspace{4mm}
}
\end{center}

\begin{center}
{ Evgueni Alexeev$^\dagger$ and Raphael Flauger$^\dagger$ }
\end{center}

\begin{center}
\vskip .6cm

\textsl{$^\dagger$ Department of Physics, UC San Diego, La Jolla, CA 92093, USA}

\end{center}

\vspace{.5cm}
{\noindent\textbf{Abstract} \\[0.1cm]
\hrule \vspace{0.3cm}
\noindent
In a cosmological setting, particle production is ubiquitous. It may occur as a consequence of the expansion of the background or because a field couples to other degrees of freedom that evolve with time. The process is well understood in the context of quantum field theory, and calculable as long as the produced quanta are weakly interacting. For extended objects like strings and membranes a second quantized formulation is much less developed than for particles. In this light, we revisit particle production from a first quantized perspective. We show how to obtain occupation numbers, both from the vacuum persistence amplitude and from the Green's function. We also derive the much less studied but phenomenologically interesting two-particle wavefunction of the produced quanta.}
\vspace{0.3cm}
 \hrule

\end{titlepage}
{
  \hypersetup{linkcolor=black,linktoc=all}
  \tableofcontents
}
\vspace{0.3cm}

\baselineskip = 16pt

\pagebreak
\section{Introduction and Summary}

A Poincar\'e-invariant state is necessarily also an eigenstate of the Hamiltonian. As a consequence, in a Poincar\'e invariant theory there can be no particle production unless quanta are already present in the initial state. As soon as Poincar\'e invariance is broken particles can in general be produced even if there are none at early times. The most famous example of particle production is Schwinger pair production in an electric field~\cite{Sauter:1931zz,Schwinger:1951nm}. In cosmology, particle production is ubiquitous. It may occur simply as a consequence of the expansion of the background~\cite{Parker:1968mv,Parker:1969au}. (See~\cite{Ford:2021syk} for a recent review.) For example, a scalar field with mass $m$ in an expanding background can be related to a scalar field in Minkowski space with an effective time-dependent mass\footnote{See Appendix~\ref{app:mass_expansion} for a derivation.}
\begin{equation}
  m^2_{\rm eff}(t) = a^2(t) \left(m^2+\left(\xi-\frac16\right)R\right)\,,
\end{equation}
where $a(t)$ is the scale factor, and $\xi$ is the non-minimal coupling of the scalar field to the Ricci scalar of the expanding spacetime, $R$. The time dependence of the effective mass can lead to particle production even if the field is initially in its adiabatic ground state.

In addition, particle production may occur because a field couples to other degrees of freedom that evolve with time and the field acquires a time-dependent mass. Examples include the process of reheating after inflation~\cite{Dolgov:1989us,Kofman:1994rk,Kofman:1997yn,Amin:2014eta,Allahverdi:2010xz}, phase transitions~\cite{Traschen:1990sw}, or non-trivial field dynamics during inflation~\cite{Kofman:2004yc,Barnaby:2009mc,Green:2009ds,Flauger:2016idt}.

As a concrete example of this type, we can consider a simple two-field model whose interaction Lagrange density is given by 
\begin{equation}
\mathcal{L}_{\rm int} = -\frac{1}{2}g^2\phi^2\chi^2 \,.\label{chi_coupling}
\end{equation}
If the field $\phi$ evolves with time, the masses of the $\chi$-fields evolve with time, and particle production will generally occur.

In the context of quantum field theory, particle production is well understood and readily calculable as long as there is a notion of $in$- and $out$-states, and the produced particles are weakly interacting. The standard approach is to compute the Bogoliubov coefficients that relate the creation and annihilation operators in the $in$ and $out$ regions~\cite{Parker:1969au,Birrell:1982ix}. (See, e.g.,~\cite{Ford:2021syk} for a review.)

Inflation~\cite{Guth:1980zm,Linde:1981mu,Albrecht:1982wi} is sensitive to physics at the Planck scale~\cite{Easther:2001fi}. This motivates considering inflation in a framework in which Planck-suppressed corrections are at least in principle calculable, such as string theory. In addition to fundamental strings, string theory contains other extended objects such as $D$-branes~\cite{Polchinski:1995mt}, and it is natural to consider the production of extended objects~\cite{Gubser:2003hm,Gubser:2003vk,Friess:2004zk,Senatore:2011sp}. As for particles, the time dependence may occur because the geometry evolves with time, or because the tension depends on fields that evolve with time. One example would be a modulus that controls the size of the cycle wrapped by a brane~\cite{Gubser:2003hm,Gubser:2003vk}. 

For strings and membranes, a second-quantized formulation is much less developed, and it is natural to revisit particle production from a first-quantized perspective in order to generalize the lessons learned for the case of particles to strings and membranes~\cite{Hamilton:2003xr,Silverstein:2014yza}.  

To make contact between the first- and second-quantized descriptions, we work with quantities that can readily be computed in either formulation. The first quantity we consider is the vacuum persistence amplitude, or equivalently the effective action $W$. These are related by 
\begin{equation}
    \langle 0,\mathrm{out} | 0, \mathrm{in}\rangle = e^{iW} \,.
\end{equation}
A convenient form for the effective action in the first-quantized formulation for a particle with a time-dependent mass is
\begin{equation}
  i W = \frac 12  \int d^3 x \int \frac{d^3p}{(2\pi)^3}\int_0^\infty\frac{dT}{T} \int d x^0\langle x^0 | \exp\left( T H_p\right) | x^0\rangle \,,
\end{equation}
where the Hamiltonian is given by
\begin{equation}
  H_p = \frac12 p_0^2-\frac12 E_p(x^0)^2\,,
\end{equation} 
with $E_p(x^0)=p^2+m^2(x^0)$ is the energy of the particle. 

For general time-dependence of the mass, there may be several production events. As we will show, the leading exponential behavior of the imaginary part of the effective action associated with the $i$-th production event is controlled by the instanton action
\begin{equation}
  S_{\rm inst, i} = 2\int_{-x^4_i}^{x^4_i} dx^4 E_p(x^0_i+i x^4) \,,
\end{equation} 
where $\pm x^4_i$ are the imaginary parts of a complex conjugate pair of zeros of the energy $E_p(x^0)$ at $x^0_i \pm i x^4_i$.
For positive definite masses that vary sufficiently slowly, the imaginary part of the effective action can be evaluated explicitly and is given by the simple expression
\begin{equation}
  \mathrm{Im} W = \frac{1}{4}\int d^3 x\int\frac{d^3p}{(2\pi)^3}\sum_i \ln\left(1+e^{-S_{\rm inst, i}}\right) \,.
\end{equation}

The vacuum persistence amplitude contains all the desired information about occupation numbers and number densities, which has been the focus of most previous work. A quantity that is conceptually and phenomenologically interesting but much less studied is the two-particle wavefunction of the produced pairs. The second quantity we consider is the Green's function for the scalar field, because its positive frequency part encodes the two-particle wavefunction. A convenient way to represent the Green's function in a first quantized form in a cosmological setting is
\begin{equation}
  G(x_f,x_i) = \frac12\int_0^\infty dT \int \frac{d^3 p}{(2\pi)^3} e^{i\mathbf{p}\cdot(\mathbf{x}_f-\mathbf{x}_i)}\langle x_f^0|\exp\left(T H_p\right)|x_i^0\rangle\,.
\end{equation}
We evaluate this expression by inserting a complete set of energy eigenstates and show that long after a single production event the two-particle wavefunction is given by 
\begin{equation}
  \Psi(\mathbf{x}_1,\mathbf{x}_2, x^0) = \mathcal{N}\int \frac{d^3 p}{(2\pi)^3} e^{i\mathbf{p}\cdot(\mathbf{x}_1-\mathbf{x}_2)}\frac{e^{-S_{\rm inst}/2}}{2E_p(x^0)} \exp\left(-2i\int^{x^0}_{x^0_1} d x\, E_p(x)\right) \,. 
\end{equation}
We also give an exact relation between the occupation numbers and the reflection coefficient for the scattering of a particle from the potential $-E^2_p(x^0)/2$ so that occupation numbers can readily be obtained for any mass whose time-dependence corresponds to a quantum mechanical potential for which the reflection coefficient is known.

This paper is organized as follows. As a warm-up, we review pair production for particles with a time-dependent mass in quantum field theory in Section~\hyperref[sec:pp]{\S 2}. We then introduce the vacuum persistence amplitude and evaluate the leading exponential and the 1-loop prefactor for slowly varying masses in two different ways in Section~\hyperref[sec:pp]{\S 3}. In Section~\hyperref[sec:propagator]{\S 4} we derive the 2-particle wavefunction in quantum field theory and use it to find the typical separation between the particles. Finally, in Section~\hyperref[sec:worldline_prop]{\S 5} we study the Green's function and two-particle wavefunction in first quantization, both in the Hamiltonian and path integral formulation. 

Our paper contains several appendices. We show how particle production in an expanding universe is related to particle production from a time-dependent mass in Appendix~\ref{app:mass_expansion}. Appendix~\ref{app:exact} provides a review of the exact solution of the mode equation for a particle whose mass-squared exhibits quadratic time dependence. In Appendix~\ref{app:propagator}, we provide a detailed discussion of the Green's function for a particle with quadratic mass-squared. We give derivations in the Hamiltonian and path integral formalism as well as in quantum field theory and show explicitly how the three are related. In Appendix~\ref{app:ratio}, we give the details of the evaluation of a functional determinant needed in the derivation of our main result in Section~\hyperref[sec:pp]{\S 3} for the effective action. Finally, in Appendix~\ref{app:time}, we include a comparison between an analytic approximation of the wavefunction at the time of production to a numerical evaluation. 

\section{Occupation numbers in quantum field theory}\label{ssec:FTreview}

Let us consider a scalar field with an effective time-dependent mass $m(t)$, caused either by the expansion of the universe (see Appendix~\ref{app:mass_expansion}) or by its interactions with degrees of freedom that evolve with time. We can expand the field in terms of time-dependent mode functions 
\begin{equation}
\chi(\mathbf{x},t)  = \int \frac{d^3 p}{(2\pi)^3}  \left[a_{\mathbf{p}}u_p(t) e^{i\mathbf{p}\cdot\mathbf{x}}+a^\dagger_{\mathbf{p}}u_p^*(t)e^{-i\mathbf{p}\cdot\mathbf{x}} \right]\label{eq:chi_fourier}\,,
\end{equation}
where $\mathbf{p}$ is the Fourier momentum, $p=|\mathbf{p}|$, and $a_{\mathbf{p}}$ and $a^\dagger_{\mathbf{p}}$ are the annihilation and creation operators for the mode with Fourier momentum $\mathbf{p}$, respectively, and the mode functions obey\footnote{We assume that the mode functions are normalized so that the creation-annihilation operators satisfy the commutation relation $\left[a(\mathbf{p}),a^{\dagger}(\mathbf{p}')\right]=(2\pi)^3\delta(\mathbf{p}-\mathbf{p}')$.}
\begin{equation}
\ddot{u}_p(t) + \left(p^2 + m(t)^2\right) u_p(t)=0\,.\label{eq1}
\end{equation}
Assuming that the mode evolution is adiabatic long before and long after any production events, it is natural to consider mode functions that correspond to the positive frequency solution at early times, often denoted $u_p^{\rm in}(t)$, and mode functions that correspond to the positive frequency solution at late times, often denoted $u_p^{\rm out}(t)$, along with the associated annihilation operators, $a_{\mathbf{p}}^{\rm in}$ and $a_{\mathbf{p}}^{\rm out}$. To be explicit, in the WKB limit we have
\begin{equation}
u_p^{\rm in}(t)\propto \frac{1}{\sqrt{2 E_p(t)}}\exp\left[-i\int^t E_p(t)\right]\qquad\text{as}\qquad t\to-\infty\,,
\end{equation}
where $E_p(t) = \sqrt{p^2 + m(t)^2}$, and similarly for $u_p^{\rm out}(t)$ as $t\to\infty$.
Since the most general solution of equation~\eqref{eq1} can be expressed as a linear combination of $u_p^{\rm out}(t)$ and its complex conjugate, we can also express $u_p^{\rm in}(t)$ as such a linear combination
\begin{equation}
u_p^{\rm in}(t) = \alpha_p u_p^{\rm out}(t) + \beta_p^*u^{* \rm out}_p(t)\,,\label{eq:in_out}
\end{equation}
and we find that the $in$- and $out$-annihilation operators are related by the Bogoliubov transformation
\begin{equation} 
a_{\mathbf{p}}^{\rm out} = \alpha_p a_{\mathbf{p}}^{\rm in} + \beta_p a^{\dagger \rm in}_{-{\mathbf{p}}} \label{bogo}
\end{equation}
To ensure that the creation and annihilation operators satisfy the usual commutation relations, the Bogoliubov coefficients must satisfy $|\alpha_p|^2 - |\beta_p|^2 = 1$.

We can compute the expected number of particles with momentum ${\mathbf{p}}$ as seen by a late-time observer in the state that the early-time observer experienced as the vacuum, $\ket{0,\rm in}$, defined by $a_{\mathbf{p}}^{\rm in}\ket{0,\rm in}=0$. This is given by
\begin{align}
\bra{0,\rm in}N_{\mathbf{p}}^{\rm out}\ket{0,\rm in} &= \bra{0,\rm in}a_{\mathbf{p}}^{\dagger \rm out}a_{\mathbf{p}}^{\rm out}\ket{0,\rm in} \nonumber\\
&=\bra{0,\rm in}(\beta_p^* a^{\rm in}_{-\mathbf{p}})(\beta_p a^{\dagger \rm in}_{-\mathbf{p}})\ket{0,\rm in} = \int d^3x\; |\beta_p|^2 \,.
\end{align}
So we see that particle production is captured by the Bogoliubov coefficient $\beta_p$.

A quantity that contains this information and can be computed conveniently in both quantum field theory and the worldline formalism is the vacuum persistence amplitude
\begin{equation}
\langle 0,\mathrm{out} | 0, \mathrm{in}\rangle = \mathcal{N}e^{iW}\,,
\end{equation}
where $\mathcal{N}$ is a normalization constant. 
If particle production occurs, $W$ will develop a positive imaginary part, and with appropriate normalization, the probability that the system will remain in the vacuum is given by
\begin{equation}
|\langle 0,\mathrm{out} | 0, \mathrm{in}\rangle|^2 = 1 - P(\text{particle production}) = \ e^{-2\mathrm{Im} W}\,.
\end{equation}
If the imaginary part is small and the production is dominated by the production of pairs of particles we can write
\begin{equation}
P(\mathrm{pairs}) \approx 2\mathrm{Im}\, W \label{pair_probability}\,.
\end{equation}
This can be expressed in terms of the Bogoliubov coefficients. By using the relations between the $in$ and $out$ creation/annihilation operators in equation~\labelcref{bogo}, the $in$-vacuum can be expressed as a superposition of $out$-particle states
\begin{equation}
\ket{0,\rm in} = e^{i\varphi}\prod_{\mathbf{p}} \frac{1}{|\alpha_p|^{1/2}} \exp \left(\frac{\beta_p}{2\alpha_p^*} a^{\dagger \rm out}_{\mathbf{p}} a^{\dagger \rm out}_{-\mathbf{p}} \right) \ket{0,\rm out} \,,
\end{equation}
where $\varphi$ is an overall phase, and the product should be taken over half of momentum space if the particles are their own anti-particles and the full momentum space if they carry any charges. The vacuum persistence amplitude is
\begin{equation}
\braket{0,\rm out | 0,\rm in} = e^{i\varphi}\prod_{\mathbf{p}} \frac{1}{|\alpha_p|^{1/2}} \,.
\end{equation}
Since the Bogoliubov coefficients satisfy $|\alpha_p|^2-|\beta_p|^2 = 1$, we can write
\begin{equation}
|\langle 0,\mathrm{out} | 0, \mathrm{in}\rangle|^2 = \prod_{\mathbf{p}} |\alpha_p|^{-1} = \exp\left(-\frac{1}{2}\int d^3 x \int \frac{d^3 p}{(2\pi)^3}\log(1+|\beta_p|^2)\right)\,, \label{1.12}
\end{equation}
where we have assumed that the particles are their own antiparticles. For small $|\beta_p|$, we can use~\labelcref{pair_probability} to derive the expected number density of pairs
\begin{equation}
\braket{n_\chi} \approx \frac{2 \mathrm{Im}\, W}{\mathrm{Vol}} = \frac{1}{2}\int \frac{d^3 p}{(2\pi)^3}|\beta_p|^2\,. \label{n_chi_eqn}
\end{equation}

So far this was general. As an example, let us consider the scenario in which the inflaton slowly rolls down its potential and one of the degrees of freedom it couples to momentarily becomes light. Assuming that production occurs on a timescale that is much shorter than the underlying $\phi$ dynamics, we can take $\dot{\phi}\approx \text{const}$, so that the mass near the production event is well approximated by 
\begin{equation}
m(t)^2 = m_0^2 + g^2\dot{\phi}^2 t^2\,.~\label{eq:moft}
\end{equation}
The general solution to~\eqref{eq1} is then given in terms of parabolic
cylinder functions $D_\nu(z)$ and the Bogoliubov coefficient $\beta_p$ is given by (see Appendix~\ref{app:exact} for details.)
\begin{equation}
\beta_p = i\exp \left(\frac{-\pi (m_0^2+p^2)}{2 g\dot{\phi}}\right) \,.\label{bk}
\end{equation}
So the occupation number $n({\mathbf{p}})$ for particles with momentum ${\mathbf{p}}$ is
\begin{equation}
n({\mathbf{p}}) = \exp\left(\frac{-\pi(m_0^2 + p^2)}{g\dot{\phi}}\right) \label{Nk_nk}\,.
\end{equation}
For $\beta_p$ given by \labelcref{bk}, we can evaluate the right-hand side of \labelcref{n_chi_eqn} by integrating over the momenta to obtain \\
\begin{equation}
\langle n_\chi \rangle \approx \frac{1}{2}\int \frac{d^3 p}{(2\pi)^3} \exp\left(\frac{-\pi(m_0^2 + p^2)}{g\dot{\phi}}\right) = \frac{(g\dot{\phi})^{3/2}}{2(2\pi)^3}e^{-\frac{\pi m_0^2}{g\dot{\phi}}} \,.\label{oc}
\end{equation} 
Note that the Boltzmann factor $e^{-\pi m_0^2 / g\dot{\phi}}$ is non-perturbative in the coupling $g$. The relationship between the number of pairs produced and the imaginary part of the effective action $W$ will hold in general, irrespective of how $W$ is computed. The field theory description is convenient in the case of particle production, but is much less developed for strings and other extended objects. So we now derive the effective action from the vacuum persistence amplitude in a first-quantized description that generalize to extended objects.

\section{Occupation numbers in first quantization}\label{sec:pp}

The vacuum persistence amplitude for a real scalar field can be expressed in terms of the determinant of the Klein-Gordon operator $\Delta$ as
\begin{equation}
\langle 0,\mathrm{out} | 0, \mathrm{in}\rangle = e^{iW}= \mathcal{N}(\mathrm{det}'\Delta)^{-1/2}\,,
\end{equation}
where $\mathcal{N}$ is a normalization constant and the prime on the determinant indicates that any zero modes have been removed. Using the identity $(\mathrm{det}'\Delta)^{-1/2} = \exp(-\frac{1}{2}\mathrm{Tr}'\log \Delta)$ and introducing a Schwinger parameter $T$, we can write the effective action formally as\footnote{As it stands, this is of course divergent in the UV where $T\to 0 $. We implicitly assume that the theory is regulated and renormalized as usual. We will employ $\zeta$ function regularization, but other choices like dimensional regularization or Pauli-Villars are, of course, just as good.}  
\begin{equation}
\langle 0,\mathrm{out} | 0, \mathrm{in}\rangle =\tilde{\mathcal{N}}\exp\left(\frac{1}{2} \int d^4 x \lim_{y^\mu\rightarrow x^\mu} \int_0^\infty \frac{dT}{T} \, \bra{x^\mu}e^{-\frac{1}{2}T\Delta}\ket{y^\mu}\right)\,. \label{1.16}
\end{equation}
Interchanging the order of integration, this leads to the following form for the effective action 
\begin{equation}
  i W = \frac 12 \int_0^\infty \frac{dT}{T} \int d^4x\langle x^\mu | \exp\left(- T H\right) | x^\mu\rangle \,,\label{eq:partition_h}
\end{equation}
where for a particle with spacetime-dependent mass, the Hamiltonian is given by
\begin{equation}
H = \frac12 p_\mu p^\mu + \frac12 m^2(x)\,.\label{eq:hamiltonian}
\end{equation}
Of course, this can equivalently be written as a Euclidean path integral for a particle moving in four dimensions\footnote{We could equivalently work with a real time path integral by inserting an $i$ in the exponent of equation~\eqref{1.16}. This has advantages and disadvantages, but we felt that the advantages of the Euclidean path integral outweighed the disadvantages. We are most interested in the imaginary part of the effective action. One advantage of the Euclidean formulation is that factors of $i$ are much easier to track. An advantage of the real time path integral makes is that it is clearer how to deal with poles that arise on the real $T$-axis in the Euclidean formulation.}
\begin{equation}
iW = \frac{1}{2} \int_0^\infty \frac{dT}{T} \oint \mathcal{D} x^\mu \exp\left[-\int_0^1 d\tau \, \left(\frac{1}{2T}\dot{x}_\mu \dot{x}^\mu + \frac{T}{2}m(x)^2\right) \right]\label{eq:partition_path}
\end{equation}
So far this holds for a mass with arbitrary dependence on the spacetime coordinates $x^\mu$. If we assume that it is only a function of time, we can treat the spatial coordinates exactly and simplify the problem. One option is to simply perform the path integral. However, for our purposes, it is more convenient to insert a complete set of momentum eigenstates into~\eqref{eq:partition_h}. This leads to
\begin{equation}
  i W = \frac 12  \int d^3 x \int \frac{d^3p}{(2\pi)^3}\int_0^\infty\frac{dT}{T} \int d x^0\langle x^0 | \exp\left( T H_p\right) | x^0\rangle \,,\label{eq:partition_p}
\end{equation}
with the Hamiltonian
\begin{equation}
  H_p = \frac12 p_0^2-\frac12 E_p(x^0)^2\,,\label{eq:hamiltonian_p}
\end{equation} 
where $E_p(x^0)^2=m(x^0)^2+p^2$. We can think of this as the Hamiltonian of a 1-dimensional quantum mechanical system with potential $-E_p(x^0)^2/2$, and we see that we can find the occupation numbers whenever the Green's function for the potential $-m(x^0)^2/2$ is known and can be integrated. Inserting a complete set of energy eigenstates, we see that we can also obtain the occupation numbers if the energy eigenvalues of any bound states and the density of states for the continuum are known.

Since the effective action involves an integral over $x^0$, we need control over the Green's function for all $x^0$. As we will discuss later, the occupation numbers can also be obtained from the late time behavior of the Green's function. As long as the energy is at least positive definite asymptotically for large $|x^0|$, they can be expressed in terms of the reflection coefficient for the potential $-E_p(x^0)^2/2$ associated with the energy eigenfunction with eigenvalue zero, which are often easier to compute than the full Green's function. So this description readily allows us to compute occupation numbers for any potential for which the reflection coefficient is already known or can be computed.

Of course, expression~\eqref{eq:partition_p} can also be represented in the form of a path integral 
\begin{equation}
  iW = \frac12\int d^3 x \int \frac{d^3p}{(2\pi)^3}\int_0^\infty \frac{dT}{T}  \oint \mathcal{D} x^0 \exp\left[-\int_0^1 d\tau \, \left(-\frac{1}{2T}(\dot{x}^0)^2 + \frac{T}{2}E_p(x^0)^2\right) \right]\,.\label{eq:worldline_p}
  \end{equation}
So far, we have not yet defined the measure for our path integrals in equations~\eqref{eq:partition_path} and~\eqref{eq:worldline_p}. However, comparing equations~\eqref{eq:partition_path} and equations~\eqref{eq:partition_p} and~\eqref{eq:hamiltonian_p} we see that we must define it so that 
\begin{equation}
    \int_x^x \mathcal{D} x \exp\left[- \int_0^1 d\tau \frac{1}{2T}\dot{x}^2 \right] = \int_{-\infty}^{\infty} \frac{dp}{2\pi} e^{-\frac{1}{2} T p^2}= \frac{1}{\sqrt{2\pi T}}\,. 
\end{equation}

Depending on the problem of interest, it may be more convenient to use the Hamiltonian or the path integral representation. The Hamiltonian formulation is much easier to work with for masses with a time dependence that involves sudden changes that one might want to approximate as a step or a series of steps. The path integral representation is most convenient in the limit in which the action for the new path is large and we can work in the semi-classical approximation.

Irrespective of the representation we chose, we see that the functional form of $E_p(x^0)$, which is guaranteed by the Lorentz invariance of the system on short timescales, implies that it is sufficient to obtain the solution for $\mathbf{p}=0$. The dependence on spatial momenta can be restored by replacing the mass in by the energy $m_0^2\to m_0^2+p^2$ in the final result. 

From our discussion, it is clear that there is nothing special about the assumption that the mass is only a function of time rather than a single spatial direction, and we could follow the equivalent steps for a mass that depends only on a single spatial direction. So while we are interested in the production of particles in a cosmological setting, the methods we discuss are not limited to a cosmological setting and can also be applied to compute particle production in external fields with large spatial gradients.  

\subsection{Path integral for slowly varying mass}\label{sec:slowmass}
The most common way to represent the trace in~\eqref{eq:partition_p} as a path integral is as a sum over all periodic paths. We turn to this in subsection~\ref{ssec:saddle}, but for now let us take the expression at face value and sum over all paths that start and end at some fixed $x^0$ and then integrate over $x^0$
\begin{equation}
  iW = \frac{V}{2} \int \frac{d^3p}{(2\pi)^3}\int_0^\infty \frac{dT}{T}  \int dx^0 \int_{x^0}^{x^0} \mathcal{D} x^0 \exp\left[-\int_0^1 d\tau \, \left(-\frac{1}{2T}(\dot{x}^0)^2 + \frac{T}{2}E_p(x^0)^2\right) \right]\,.\label{eq:worldline_x0}
\end{equation}
This has the advantage that our calculations directly tie to those for the propagator, which we will consider later, and that the fluctuations around the classical path obey Dirichlet boundary conditions for which the evaluation of the fluctuation determinant is somewhat easier. 

Let us assume that we have found a classical path $\bar{x}^0(\tau)$ that starts and ends at $x^0$, i.e. a solution of the equation of motion
\begin{equation}
  \ddot{\bar{x}}^0 + \frac{T^2}{2}\frac{d m^2}{d\bar{x}^0} = 0\,,\label{eq:worldline_x0_eom}
\end{equation}
with the boundary conditions $\bar{x}^0(0)=\bar{x}^0(1)=x^0$. To see that such paths exist at least for sufficiently small $T$, note that if the derivative of the potential at $x^0$ is non-zero, there will be a solution in which the particle rolls up the hill and then back down. If the derivative vanishes, there will be a solution in which the particle stays in place. Since the $T$ integrand will be an analytic function, finding it even on a small open set is sufficient to find it for all $T$.  Writing the general path as $x^0(\tau)=\bar{x}^0(\tau)+\delta x^0(\tau)$, the effective action becomes
\begin{multline}
  iW = \frac{V}{2} \int \frac{d^3p}{(2\pi)^3}\int_0^\infty \frac{dT}{T}  \int dx^0 \exp\left[-\int_0^1 d\tau \left(-\frac{1}{2T}(\dot{\bar{x}}^0)^2 + \frac{T}{2} E_p^2(\bar{x}^0)\right) \right]\\\times\int \mathcal{D} \delta x^0 \exp\left[-\int_0^1 d\tau \, \left(-\frac{1}{2T}(\delta\dot{x}^0)^2 + \frac{T}{2}\left(m^2(\bar{x}^0+\delta x^0)-m^2(\bar{x}^0)-\frac{d m^2}{d\bar{x}^0}\delta x^0\right)\right) \right]\,.\label{eq:worldline_x0_exp}
\end{multline}
So far this is exact. Let us assume that the worldline theory is weakly coupled around the production event so that it is a good approximation to work to second order in fluctuations. This will be the case if the mass is sufficiently slowly varying, specifically that $dm^2/dx^{n} \ll (dm^2/dx^{2})^{(n+2)/4}$ for $n>2$ around the production event. This leads to the effective action 
\begin{multline}
  iW = \frac{V}{2} \int \frac{d^3p}{(2\pi)^3}\int_0^\infty \frac{dT}{T}  \int dx^0 \exp\left[-\int_0^1 d\tau \left(-\frac{1}{2T}(\dot{\bar{x}}^0)^2 + \frac{T}{2} E_p^2(\bar{x}^0)\right) \right]\\\times\int \mathcal{D} \delta x^0 \exp\left[-\int_0^1 d\tau \, \left(-\frac{1}{2T}(\delta\dot{x}^0)^2 + \frac{T}{4} \frac{d^2 m^2}{d\bar{x}^{0^2}}(\delta x^0)^2\right) \right]\,.\label{eq:worldline_x0_quad} 
\end{multline}
and performing the Gaussian integral
\begin{multline}
  iW = i \frac{V}{2} \int \frac{d^3p}{(2\pi)^3}\int_0^\infty \frac{dT}{T} \frac{1}{\sqrt{2\pi T}} \\\times\int dx^0 \exp\left[-\int_0^1 d\tau \left(-\frac{1}{2T}(\dot{\bar{x}}^0)^2 + \frac{T}{2} E_p^2(\bar{x}^0)\right) \right]\sqrt{\frac{\det \left(-\frac{1}{T^2}\frac{\partial}{\partial \tau^2}\right)}{\det \left(-\frac{1}{T^2}\frac{\partial}{\partial\tau^2} - \omega(\tau)^2\right)}}\,\label{eq:worldline_x0_det} 
\end{multline}
where 
\begin{equation}
  \omega(\tau)^2 = \frac{1}{2} \left.\frac{d^2 m^2}{dx^{0^2}}\right|_{\bar{x}^0(\tau)}\,.
\end{equation} 
It is instructive to evaluate the determinant using the method by Gelfand and Yaglom~\cite{Gelfand:1959nq,Coleman:1985rnk,Dunne:2007rt}, which reduces the evaluation of the ratio of functional determinants to finding the solution of the equation
\begin{equation}
  \ddot{\tilde{x}}+ T^2\omega^2(x^0)\tilde{x} = 0\,,\label{eq:worldline_fluc}
\end{equation}
with initial conditions $\tilde{x}(0)=0$ and $\dot{\tilde{x}}(0)=1$. 
For slowly varying background fields, we can find the solution in the WKB approximation 
\begin{equation}
  \tilde{x}(\tau) = \frac{\sin(T\Omega(\tau))}{T\sqrt{\omega(0)\omega(\tau)}}\,,
\end{equation}
where $\Omega(\tau) = \int_0^\tau d\tau' \omega(\tau)$. The ratio of functional determinants is then given by
\begin{equation}
  \frac{\det \left(-\frac{1}{T^2}\frac{\partial}{\partial \tau^2}\right)}{\det \left(-\frac{1}{T^2}\frac{\partial}{\partial\tau^2} - \omega(\tau)^2\right)} = \frac{\omega(0)T}{\sin(T\Omega(1))}\label{eq:det_ratio}\,,
\end{equation}
where we have used $\omega(1)=\omega(0)$. For very small $T$, we know that all eigenvalues are positive, and we see from our result~\eqref{eq:det_ratio} that this remains true as long as $T\Omega(1)< \pi$. For this range, the contribution to the effective action is real. As $T$ increases, we encounter a pole in the ratio of functional determinants at $T\Omega(1)=\pi$ where the first eigenvalue crosses zero. From the solution~\eqref{eq:det_ratio}, we see that the operator has a single negative eigenvalue until we reach the pole at $T\Omega(1)=2\pi$, where the next eigenvalue crosses zero, and so on.

To make further progress, we now need the background solution $\bar{x}^0(\tau)$ to evaluate the remaining part of the integrand. By taking the derivative of equation~\eqref{eq:worldline_x0_eom}, we see that $\dot{\bar{x}}^0(\tau)$ also obeys equation~\eqref{eq:worldline_fluc}. For a path that starts and ends at $x^0$ at $\tau=0$ and $\tau=1$, the velocity must vanish at $\tau=1/2$. So the solution in the WKB approximation must be of the form
\begin{equation}
  \dot{\bar{x}}(\tau) = \frac{A}{\sqrt{\omega(\tau)}}\sin(T\Omega_{1/2}-T\Omega(\tau))\,,
\end{equation}
where $\Omega_{1/2}=\Omega(1/2)=\Omega(1)/2$. 
Had we worked with periodic boundary conditions, this would be a zero mode of the fluctuation determinant corresponding to translations of the background solution in $\tau$. Here it is not in the spectrum of fluctuations unless $\dot{\bar{x}}^0(0)=\dot{\bar{x}}^0(1)=0$, which occurs for the first time at $T\Omega_{1/2}=\pi$, or equivalently at $T\Omega(1)=2\pi$. The solution for the background path can then be obtained by integrating
\begin{equation}
  \bar{x}^0(\tau) = x^0 + \frac{A}{T\omega(\tau)^{3/2}}\left(\cos(T\Omega_{1/2}-T\Omega(\tau)) -\cos(T\Omega_{1/2})\right)\,.\label{eq:background_wkb}
\end{equation}
The coefficient $A$ can be determined in terms of the turning point $x_{1/2}^0$ of the solution by using energy conservation. It is given by 
\begin{equation}
  A= \pm T\sqrt{m^2(x_{1/2})-m^2(x^0)}\frac{\sqrt{\omega(0)}}{\sin(T\Omega_{1/2})}\,,
\end{equation}
with the sign matching the sign of $dm^2/dx^0$.

The location of the turning point is itself a function of $T$, and from~\eqref{eq:background_wkb} we see that it is given by the solution of
\begin{equation}
x^0_{1/2} = x^0 \pm \sqrt{m^2(x_{1/2})-m^2(x^0)}\frac{\omega(0)^{1/2}}{\omega(\frac12)^{3/2}}\tan\left(\frac{T\Omega_{1/2}}{2}\right)\,.
\end{equation}
If the mass is slowly varying, we can work in a derivative expansion in $\omega(x^0)$ just like we would in a heat kernel expansion. At leading order, this amounts to neglecting derivatives of $\omega(x^0)$. In practice, when evaluating the integrand at $x^0$ this amounts to working with
\begin{equation}
  m^2(x)=m^2(x^0)+m^{2\,\prime}(x^0)(x-x^0)+\omega^2(x^0)(x-x^0)^2\,,
\end{equation}
keeping the $x^0$-dependence of $\omega(x^0)$ but neglecting derivatives acting on it. In this approximation, we have $\omega(\tau)=\omega(x^0)$ independent of $\tau$ and hence $\Omega_{1/2}=\omega(x^0)/2$, and we can explicitly find the turning point as a function of $T$. The background solution for the path starting and ending at $x^0$ is then given by
\begin{equation}
\bar{x}^0(\tau)=x^0-\frac{m^{2\,\prime}(x^0)}{2\omega^2(x^0)}+\frac{m^{2\,\prime}(x^0)}{2\omega^2(x^0)}\frac{\cos(T\omega(x^0)/2-T\omega(x^0)\tau)}{\cos(T\omega(x^0)/2)}\,.
\end{equation}
We will need it to compute the energy as well as the action. So let us also give the expression for the velocity
\begin{equation}
\dot{\bar{x}}^0(\tau) = \frac{m^{2\,\prime}(x^0) T}{2\omega(x^0)}\frac{\sin(T\omega(x^0)/2-T\omega(x^0)\tau)}{\cos(T\omega(x^0)/2)}\,.
\end{equation}
The energy associated with this motion is given by 
\begin{equation}
  \mathcal{E}= \frac12 E^2_p(x^0) + \frac12\left(\frac{m^{2\,\prime}(x^0)}{2\omega^2(x^0)}\right)^2\tan^2(T\omega(x^0)/2)\,,
\end{equation}
and the action evaluated on this path is given by
\begin{eqnarray}
  S[\bar{x}^0]&=&\int_0^1 d\tau \left[-\frac{1}{2T}\dot{\bar{x}}^0(\tau)^2 +\frac{T}{2} E^2_p(\bar{x}^0(\tau))\right] \\
  &=& \frac{T}{2}\left(E_p^2(x^0)-\left(\frac{m^{2\,\prime}(x^0)}{2 \omega(x^0)}\right)^2\right)+\left(\frac{m^{2\,\prime}(x^0)}{2 \omega(x^0)}\right)^2\frac{\tan(T\omega(x^0)/2)}{\omega(x^0)}\,.
\end{eqnarray}
Notice that  the energy and action for the path that goes up the hill diverge as $T\omega(x^0)\to \pi$, but by analyticity our results extend to the complex $T$ plane. Putting everything together, we find that the effective action is given by 
\begin{multline}
  W = \frac{V}{2} \int \frac{d^3p}{(2\pi)^3}\int_0^\infty \frac{dT}{T} \frac{1}{\sqrt{2\pi T}}\int dx^0 \sqrt{\frac{\omega(x^0)T}{\sin(\omega(x^0)T)}} \\\times\exp\left[-\frac{T}{2}\left(E_p^2(x^0)-\left(\frac{m^{2\,\prime}(x^0)}{2 \omega(x^0)}\right)^2\right)-\left(\frac{m^{2\,\prime}(x^0)}{2 \omega(x^0)}\right)^2\frac{\tan(T\omega(x^0)/2)}{\omega(x^0)} \right]\,.\label{eq:worldline_x0_appr} 
\end{multline}
We can evaluate the $T$ integral using the techniques described in Appendix~\eqref{app:propagator}. Before turning to this, note that if the action is large, which amounts to $E_{\mathbf{p}}^2(x^0)/\omega(x^0)\gg 1$,\footnote{Notice that $\omega(x^0)$ has mass dimension 2.} it is natural to evaluate the integral in the saddle point approximation. It is instructive to consider two different approaches.  

Let us first evaluate the integral along the real $x^0$-axis in the saddle point approximation. The saddles are given by the extrema of $m^2(x^0)$, and we will model the mass near the $i$-th saddle as 
\begin{equation}
  m^2(x^0) = m_i^2+\omega_i^2(x^0-x^0_i)^2\,.
\end{equation}
Near the saddle $\omega(x^0)\approx \omega_i$, and after shifting the integration variable, the integral we have to evaluate becomes
\begin{multline}
\int_\infty^\infty dx^0 \sqrt{\frac{\omega_i T}{\sin(\omega_i T)}}\exp\left[-\frac{T}{2}\left(p^2+m_i^2\right)-\omega_i {x^0}^2\tan(T\omega_i/2) \right]\\=\frac12\exp\left[-\frac{T}{2}\left(p^2+m_i^2\right)\right]\frac{\sqrt{2\pi T}}{\sin(\omega_i T/2)}\,. 
\end{multline}
The effective action is then given by
\begin{equation}
  W = \frac{V}{2} \int \frac{d^3p}{(2\pi)^3}\int_0^\infty \frac{dT}{T} \sum_i\frac{\exp\left(-\frac{T}{2}\left(p^2+m_i^2\right)\right)}{2\sin(\omega_i T/2)}\,.
\end{equation}
We see that the imaginary part of the effective action entirely arises from the poles of the sine function  along the real axis at $T_n=2\pi n/\omega_i$. At this stage, in the Euclidean path integral it is not immediately clear how to deal with the poles. However, from Lorentzian path integral, we see that we should pass just above each pole so that each pole contributes $-i\pi$ times the residue, half the contribution we would find had we encircled the pole. We will use this observation later. The imaginary part of the effective action is then given by 
\begin{eqnarray}
  \mathrm{Im} W &=& \frac{1}{4}\int d^3 x \int \frac{d^3p}{(2\pi)^3}\int_0^\infty \sum_i\sum_{n=1}^\infty\frac{(-1)^{n-1}}{n}\exp\left[-\frac{n\pi}{\omega_i}\left(p^2+m_i^2\right)\right]\,\nonumber\\
  &=& \frac{1}{4}\int d^3 x \int \frac{d^3p}{(2\pi)^3}\sum_i\log\left(1+\exp\left[-\frac{\pi}{\omega_i}\left(p^2+m_i^2\right)\right]\right)\,.\label{eq:ImW_slowmass}
\end{eqnarray}
As a consistency check, notice that for a single event we recover the occupation numbers we found in equation~\eqref{Nk_nk}.

It is also instructive to return to~\eqref{eq:worldline_x0_appr} and consider performing the two-dimensional integral in a saddle point approximation. The saddle points are located at $\omega(x^0)T=2\pi n$, and $E_p(x^0)=0$. Assuming our expression for the energy is positive definite for real $x^0$, we see that the saddle points must lie at complex values of $x^0$. We also note that in our earlier derivation, we encountered but ultimately narrowly avoided the poles on the real $T$ axis and never had to deal with the zero mode of the fluctuation operator explicitly. Here we see that to perform the full integral in the saddle point approximation, we must deal with the zero mode, and we will do so below. However, we can already evaluate the exponent at the saddle point without additional work and see that each saddle gives a contribution to the effective action
\begin{equation}
W_n\propto\exp\left[\frac{n\pi}{\omega(x_s)}\left(\frac{m^{2\,\prime}(x_s)}{2 \omega(x_s)}\right)^2 \right]\,, 
\end{equation}
with the saddle point $x_s$ given by the solution of $E_p(x_s)=0$. Near a minimum of the mass, we find 
\begin{equation}
  x_s=x_i^0\pm i\frac{\sqrt{p^2+m_i^2}}{\omega_i}\,,
\end{equation}
so that the effective action at the leading saddle with $n=1$ is proportional to
\begin{equation}
  W\propto\exp\left[-\frac{\pi}{\omega_i}\left(p^2+m_i^2\right)\right]\,,
\end{equation}
consistent with our earlier calculation. To determine the prefactor in this method, we have to deal with the zero mode of the fluctuation operator, and we should determine which saddle points contribute and with which factors, for example, using the techniques described in~\cite{Witten:2010cx}.

Let us now evaluate the integral over $T$ in equation~\eqref{eq:worldline_x0_appr} exactly, using the techniques described in Appendix~\eqref{app:propagator}. Performing the integral over $d$ spatial momenta and slightly rearranging~\eqref{eq:worldline_x0_appr}, we find
\begin{multline}
  W = \frac{V}{2}  \int_0^\infty \frac{dT}{T}\frac{1}{(2\pi T)^{d/2}} \int dx^0 \exp\left[-\frac{T}{2}(\mu^2-i\omega(x^0))\right] \\\times\sqrt{\frac{\omega(x^0)}{\pi}}e^{-i\pi/4}\frac{e^{-i B}}{\sqrt{1-q^2}}\exp\left[2 i B \frac{q}{1-q^2} \right]\,,
\end{multline}
where we have defined
\begin{align}
  q &= \exp(i\omega(x^0)T)\,,\\ 
  \mu^2(x^0) & = m^2(x^0)-\left(\frac{m^{2\,\prime}(x^0)}{2 \omega(x^0)}\right)^2\,,\\
  B(x^0) &=\left(\frac{m^{2\,\prime}(x^0)}{2 \omega(x^0)}\right)^2\frac{1}{\omega(x^0)}\,, 
\end{align}
We can now proceed as in Appendix~\ref{app:propagator} and apply Mehler's formula, and convert the resulting sum into a contour integral
\begin{multline}
  W = \frac{V}{2}  \int_0^\infty \frac{dT}{T}\frac{1}{(2\pi T)^{d/2}} \int dx^0 \exp\left[-\frac{T}{2}(\mu^2-i\omega(x^0))\right] \\\times -\sqrt{\frac{\omega(x^0)}{\pi}}e^{-i\pi/4}\int_\mathcal{C} \frac{dz}{2\pi i} e^{i z \omega T}\Gamma(-z)D_z\left(e^{i\pi/4} (2B)^{1/2}\right)D_z\left(-e^{i\pi/4} (2B)^{1/2}\right)\,,
\end{multline}
where the contour is the same as shown in Figure~\ref{fig:contour}. We can now perform the integral over $T$, assuming $-2<d<0$ to ensure convergence, and analytically continue the result to $d=3$. This leads to
\begin{multline}
  W = \frac{V}{2}   \int dx^0 \sqrt{\frac{\omega(x^0)}{\pi}}\left(\frac{\omega(x^0)}{2\pi}\right)^{d/2} \Gamma(-d/2) e^{i\pi (3-d)/4} \\\times \int_\mathcal{C} \frac{dz}{2\pi i} \left(z+\frac12 + i\frac{\mu^2}{2\omega(x^0)}\right)^{d/2} \Gamma(-z)D_z\left(e^{i\pi/4} (2B)^{1/2}\right)D_z\left(-e^{i\pi/4} (2B)^{1/2}\right)\,.
\end{multline}
For $d=3$, we see that the integral has resulted in a cut instead of the pole we encountered for the Green's function. We can perform the integral over $z$ by deforming the contour to run along the cut. The result can be brought into the form
\begin{multline}
  W = \frac{V}{2} \int \frac{d^3 p}{(2\pi)^3}  \int dx^0 \\\times -  e^{i\pi/4} \frac23 \frac{p^2}{\sqrt{2\omega}} \frac{\Gamma(\bar\nu+1)}{\sqrt{2\pi}}D_{-\bar\nu-1}\left(e^{i\pi/4} (2B)^{1/2}\right)D_{-\bar\nu-1}\left(-e^{i\pi/4} (2B)^{1/2}\right)\,,
\end{multline} 
where $\bar\nu = -1/2+i(\mu^2+p^2)/2\omega$. We can use the identity~\eqref{eq:Dnu_rel} to ensure the arguments of all parabolic cylinder functions have phase between $\pm 3\pi/4$. The first term yields a positive definite contribution and only affects the real part. For the purpose of particle production, we are only interested in the imaginary part which only receives contributions from the second term 
\begin{multline}
  W \supset \frac{V}{2} \int \frac{d^3 p}{(2\pi)^3}  \int dx^0 \\\times  e^{i\pi/4}  e^{i\pi\bar\nu} \frac23 \frac{p^2}{\sqrt{2\omega}} \frac{\Gamma(\bar\nu+1)}{\sqrt{2\pi}}D_{-\bar\nu-1}\left(e^{i\pi/4} (2B)^{1/2}\right)D_{-\bar\nu-1}\left(e^{i\pi/4} (2B)^{1/2}\right)\,.
\end{multline}
Since both the index $\bar\nu$ and the argument $B$ are in general functions of $x^0$, we cannot evaluate the integral for general masses.

\subsection{Path integral in the saddle point approximation}\label{ssec:saddle}

In the limit in which the action is large and the production is suppressed, it is natural to perform the integral in the saddle-point approximation. We will now consider this approach, starting with the integral over the Schwinger parameter $T$~\cite{Affleck:1981bma}. In this approximation, the effective action is given by  
\begin{equation}
iW = \frac{V}{2}\sqrt{\frac{\pi}{2}}\int\frac{d^3p}{(2\pi)^3}\,\oint{\mathcal{D}x^0} \frac{\exp\left(-S_\star[x^0]\right)}{\sqrt{S_\star[x^0] }}\,, \label{eq:worldline_saddle}
\end{equation}
where $S_\star$ is the non-local action given by 
\begin{equation}
S_\star[x^0] = \sqrt{-\int_0^1\,d\tau\,(\dot{x}^0)^2} \,\sqrt{\int_0^1\,d\tau\, E_{p}^2}\,,
\end{equation}
and the saddle point is located at
\begin{equation}
  T_\star[x^0] = \frac{\sqrt{-\int(\dot{x}^0)^2}}{\sqrt{\int E_{\mathbf{p}}^2}}\,. \label{eq:T_saddle}
\end{equation}
This approximation is valid as long as the action is large, $S_\star[x^0]\gg1$. We have inserted a factor $1/2$ in the prefactor, implicitly assuming that we will be interested in saddles along the real axis at $T_\star=2\pi n/\omega(x^0)$, which we saw contribute with a factor $1/2$. The correct prefactor might differ for any complex saddles.  

We can perform the path integral by decomposing $x^0(\tau)$ into a background solution $\bar{x}^0(\tau)$ with boundary conditions $\bar{x}^0(0)=\bar{x}^0(1)$ centered at $x^0$, and periodic fluctuations $\delta x^0$ around this background
\begin{equation}
  \int_0^1 d\tau \bar{x}^0(\tau) = x^0\,\quad\text{and}\qquad \int_0^1 d\tau \delta x^0(\tau) = 0\,.
\end{equation}
The effective action can then be written as 
\begin{equation}
  iW = \frac{V}{2}\sqrt{\frac{\pi}{2}}\int\frac{d^3p}{(2\pi)^3}\int dx^0\exp\left(-S_\star[\bar{x}^0]\right)\,\oint{\mathcal{D}\delta x^0} \frac{\exp\left(-(S_\star[\bar{x}^0+\delta x^0]-S_\star[\bar{x}^0])\right)}{\sqrt{S_\star[x^0+\delta x^0] }}\,, \label{eq:worldline_inst}
\end{equation}
So far this is exact (with appropriate definition of the measure), but it will again be most useful in cases in which expanding in fluctuations is a good approximation. To eliminate terms of first order in the perturbations, we can work with a background that obeys the equation of motion
\begin{equation}
\frac{\delta S_\star}{\delta x^0} = \frac{1}{T_\star[\bar{x}^0]}\ddot{\bar{x}}^0 + \frac12\frac{dm^2}{d\bar{x}^0}T_\star[\bar{x}^0] = 0\,.
\end{equation}
This equation implies the existence of a conserved energy, and at the saddle $T_\star$ this energy must vanish
\begin{equation}
\frac12\left(\dot{\bar{x}}^0\right)^2+\frac12 E_p^2(\bar{x}^0) T_\star[\bar{x}^0]^2 = 0\,.\label{eq:constraint}
\end{equation}
For positive definite $E_p^2(\bar{x}^0)$ and real $\bar{x}^0$ the left-hand side is positive definite, and we see that no real solutions exist. This is, of course, directly related our earlier observation that the saddle points of the two-dimensional integral correspond to points at which the energy vanishes. At these points $\bar{x}^0$ is necessarily complex (except perhaps at $\mathbf{p}=0$). This makes it natural to analytically continue $x^0$ and work with a complex coordinate $x = x^0 + i x^4$, although we will often keep working with $x^0$ to remind ourselves of the contour we are integrating over. 

Since $E_p^2(x^0)$ is real and positive definite, we know that any zeros must come in complex conjugate pairs $x=x^0_i\pm i x^4_i$, and we can interpret $x^0_i$ as the time of the production event. If the production events are well separated, we expect the dominant saddles to correspond to solutions that oscillate between the two points of a complex conjugate pair. 

We will now consider these background solutions and the corresponding instanton actions in~\ref{ssec:background} and continue our discussion of the path integral over the fluctuations in subsection~\ref{ssec:particle prefactor}.

\subsubsection{Background solution and leading exponent}\label{ssec:background}
As we discussed earlier in this section, the background solutions are (complex) solutions of the equation of motion
\begin{equation}
  \ddot{\bar{x}}^0 + \frac12\frac{dm^2}{d\bar{x}^0}T_\star[\bar{x}^0]^2 = 0\,,\label{eq:insteom}
\end{equation}
with zero energy. To determine the exponential dependence of the occupation numbers, we only need the action evaluated on this solution, not the solution itself. At zero energy, we can use equation~\eqref{eq:constraint} to write the action as 
\begin{equation}
  S_\star[\bar{x}] = T_\star\int_0^1\,d\tau\, E_p^2(\bar{x}(\tau))\,.
\end{equation}
Assuming that the production events are well separated, and considering for now the case in which $dm^2/d\bar{x}^0|_{x_i^0}=0$, the dominant solutions correspond to the motion between the two points of a complex conjugate pair $x=x^0_i\pm i x^4_i$. We can then use equation~\eqref{eq:constraint} again to write the instanton action as 
\begin{equation}
  S_{\rm inst} =T_\star\int_{-x^4_i}^{x^4_i} dx^4 \frac{d\tau}{dx^4} E_p^2(x^0_i+i x^4) + T_\star\int_{x^4_i}^{-x^4_i} dx^4 \frac{d\tau}{dx^4} E_p^2(x^0_i+i x^4)\,.
\end{equation}
Solving the constraint~\eqref{eq:constraint} for $d\tau/dx^4$, using that the velocity is positive in the first integral and negative in the second, we see that the instanton action is given by 
\begin{equation}
  S_{\rm inst} = 2\int_{-x^4_i}^{x^4_i} dx^4 E_p(x^0_i+i x^4)\label{eq:inst_contour} \,.
\end{equation}
Here we have considered the leading saddle corresponding to a single oscillation between the zeros of the energy. The action for the subleading saddles in which the particle completes $n$ oscillations is simply $n$ times the action for the leading saddle.

If $dm^2/d\bar{x}^0|_{x_i^0}\neq 0$, the velocity $dx^0/d\tau$ is necessarily non-zero at the turning point of $x^4(\tau)$ where $dx^4/d\tau=0$. From energy conservation, we see that at this point $E_p^2<0$. This implies that the corresponding solution completely encircles the two branch points, and by analyticity, the integral is still given by~\eqref{eq:inst_contour}.

While it might appear surprising that we can evaluate the action without knowing the solution, in fact, this had to be true. We know that we can write the effective action in a completely diffeomorphism invariant way by introducing an einbein. In this formulation, the einbein can be thought of a Lagrange multiplier enforcing the constraint, explaining why it played a key role in arriving at this reparametrization invariant answer. 

If needed, we can also use the constraint to find the solution. For simplicity, let us again assume that we are interested in the solution that corresponds to the motion between a complex conjugate pair of zeros of the energy, leaving the zero with positive imaginary part at $\tau=0$. In this case, $x^0(\tau)=x_i^0+i x^4(\tau)$, so that for $\tau\leq 1/2$
\begin{equation}
\tau(x_4) = \frac{1}{T_\star}\int^{x^4_i}_{x^4} \frac{dx}{ E_p(x^0_i+i x)}\label{eq:tauofx4}\,,
\end{equation}
and similarly for $\tau > 1/2$
\begin{equation}
  \tau(x_4) = \frac12+ \frac{1}{T_\star}\int^{x^4}_{-x^4_i} \frac{dx}{ E_p(x^0_i+i x)}\label{eq:tauofx4_2}\,.
  \end{equation}
To ensure the solution returns to the zero with positive imaginary part at $\tau=1$, we must have
\begin{equation}
  T_\star = 2\int^{x^4_i}_{-x^4_i} \frac{dx}{E_p(x^0_i+i x)}\,.\label{eq:Tstarint}
\end{equation}
The solution for the background path is then given by $x^0(\tau)=x^0_i+i x^4(\tau)$, where $x^4(\tau)$ is the inverse function of $\tau(x^4)$ in equation~\eqref{eq:tauofx4}.

Here we have focused on the solution that completes one period between the two zeros of the energy. The solutions that complete $n$ periods can be dealt with in the same way. The only change needed is a factor of $n$ in the numerator on the right-hand side of the equation for $T_\star$~\eqref{eq:Tstarint}. 

As an example, let us again consider a mass that has local minima near some $x^0_i$ and is well approximated by $m^2(x^0)=m^2_i+\omega^2_i(x^0-x^0_i)^2$ so that the zeros of the energy are located at $x=x_i^0\pm ix^4_i$ with $x^4_i=\sqrt{p^2+m_i^2}/\omega_i$. The instanton action associated with the $i$-th minimum is then given by 
\begin{equation}
  S_{\rm inst} = 2\int_{-x^4_i}^{x^4_i} dx_4 \sqrt{p^2+m_i^2-\omega_i^2 x_4^2}=\frac{\pi (p^2+m_i^2)}{\omega_i}\,,
\end{equation}
implying that the occupation numbers are 
\begin{equation}
  n({\mathbf{p}}) \propto e^{-S_\mathrm{inst}} = \exp\left(-\frac{\pi}{\omega_i}(p^2+m_i^2)\right)\,,
\end{equation}
in agreement with our earlier results.

To find the instanton solution, we first perform the integral~\eqref{eq:Tstarint}, which results in $T_\star=2\pi /\omega_i$. Performing the integral~\eqref{eq:tauofx4}, we see that the solution is given by
\begin{equation}
  \tau(x_4) = \frac{1}{2\pi}\left(\frac{\pi}{2}-\arcsin\left(\frac{\omega_i x_4}{\sqrt{p^2+m_i^2}}\right)\right)\,,
\end{equation}
so that 
\begin{equation}
  x(\tau) = x^0_i + i\frac{\sqrt{p^2+m_i^2}}{\omega_i}\cos\left(2\pi\tau\right)\,.
\end{equation}
There was nothing special in our choice that the solution leave the zero with positive imaginary part at $\tau=0$ so that the general solution is obtained by including a phase or equivalently replacing $\tau\to \tau-\tau_0$ in the solution above. These, of course, all have the same action, and this symmetry is responsible for the zero mode that led to the pole at $T=2\pi/\omega_i$ in section~\ref{sec:slowmass}, and that we have to take care of in the path integral over the fluctuations.

In our example, we have assumed that the behavior of the mass around the production event is well approximated by a quadratic function. To better understand the regime in which this is true, it is instructive to consider another example in which the mass is quadratic near the minimum and approaches a constant at early and late times 
\begin{equation}
m^2(x^0) = m_0^2 + \omega^2 \Delta T^2 \tanh^2\left(\frac{x^0}{\Delta T}\right)\,,
\end{equation} 
where $\Delta T$ is the timescale over which the mass changes. In this case, integrating equation~\eqref{eq:Tstarint} implies that the saddle point is now located at
\begin{equation}
T_\star = \frac{2\pi}{\omega_i}\frac{1}{\sqrt{1+(E_0/\omega\Delta T)^2}}\,,
\end{equation}
where we have defined $E_0=\sqrt{p^2+m_0^2}$. Integrating equation~\eqref{eq:tauofx4} then implies that the leading instanton solution is given by
\begin{equation}
\bar{x}^0(\tau) = i \Delta T \arctan\left(\frac{E_0}{\omega \Delta T} \frac{\cos(2\pi(\tau-\tau_0))}{\sqrt{1+(E_0/\omega\Delta T)^2 \sin^2(2\pi(\tau-\tau_0))}}\right)\,,
\end{equation}
and the instanton action is given by
\begin{equation}
S_{\rm inst}= 2 \pi \Delta T E_0 \sqrt{\left(1-\frac{1}{\sqrt{1+(E_0/\omega \Delta T)^2}}\right)\left(1-\frac{1}{1+\sqrt{1+(E_0/\omega \Delta T)^2}}\right)}\,.
\end{equation}
So if the timescale $\Delta T$ on which the mass varies is much longer than the timescale $E_0/\omega$ associated with the production, $E_0/\omega \Delta T\ll 1$, and approximating $m^2$ near the minimum as quadratic is a good approximation. This will break down for modes with high momentum, but their contribution is highly suppressed.  

Even when we cannot explicitly solve the integral~\eqref{eq:tauofx4}, we can still get a qualitative understanding for the solution by using our intuition for the motion of a particle in the potential $V(x_4)=-E^2_p(x_i^0+i x^4)T_\star^2/2$ at zero energy. To further simplify the problem, as we discussed, we can set $p=0$ and restore it at the end of the calculation.

As a final example, let us consider the production of scalar quanta in an expanding FLRW universe as an example in which the derivative of the mass at the production event is non-zero. We will consider a conformally coupled scalar in the toy geometry considered in~\cite{Ford:2021syk} for which the scale factor is given by
\begin{equation}
a^2(\eta) = \frac{1}{2}(a_f^2+a_i^2 + (a_f^2-a_i^2)\tanh(\tau/\Delta\tau))\,. \label{eq:tanh_universe}
\end{equation}
This corresponds to a scenario in which the scale factor increases from an initial value $a_i$ to a final value $a_f$ on a time scale $\Delta \tau$. 
As explained in Appendix~\ref{app:mass_expansion}, we can equivalently consider to a scalar field in Minkowski space with time-dependent effective mass 
\begin{equation}
  m_{\rm eff}^2(\tau) = a^2(\tau)m^2\,.
\end{equation}
In this case the relevant zeros of the effective mass $m_{\rm eff}^2(\tau$) correspond to
\begin{equation}
\tau_{\pm} = \Delta\tau\ln \left(\frac{a_i}{a_f}\right)\pm i \frac{\pi}{2}\Delta \tau\,.
\end{equation}
The instanton action can be obtained by integrating the energy over a contour encircling the branch points. This evaluates to
\begin{equation}
S_\mathrm{inst} = 2\pi a_i E_p\Delta \tau\,,
\end{equation}
reproducing the exponential in equation 2.40 of \cite{Ford:2021syk}.

\subsubsection{One-loop prefactor}\label{ssec:particle prefactor}
We now determine the prefactor by computing the integral over the fluctuations, expanding to second order in fluctuations around the periodic solutions $\bar{x}(\tau)$ that oscillate between points at which the energy vanishes. This will be a good approximation under the same condition on the mass as assumed in section~\ref{sec:slowmass}, and if the production events are well separated in time. We may think of this as the limit in which interactions between the instantons associated with the different production events are negligible and they can be treated as free. Keeping only the dependence on the fluctuations in the exponent, the effective action becomes  
\begin{multline}
  iW = \frac{V}{2}\sqrt{\frac{\pi}{2}}\int\frac{d^3p}{(2\pi)^3}\sum_\mathrm{saddles} \frac{\exp\left(-S_\star[\bar{x}^0]\right)}{\sqrt{S_\star[\bar{x}^0] }}\\\times\oint{\mathcal{D}\delta x^0} \exp\left[\frac12\int_0^1 \int_0^1 d\tau d\tau' T_\star \delta x^0(\tau) M(\tau,\tau') \delta x^0(\tau')\right]\,.\label{eq:worldline_saddle_quad}
\end{multline}
The measure here is defined so that
\begin{equation}
  \int {\mathcal{D}\delta x^0} \exp\left[-\frac12\int_0^1 d\tau\,  T_\star\,\delta x^0(\tau)^2\right] =1 \,, 
  \end{equation} 
and $M$ is given by 
\begin{equation}
  M(\tau,\tau') = \left(-\frac{1}{T_\star^2}\frac{\partial^2}{\partial \tau^2} - \omega^2(\tau)\right)\delta(\tau-\tau')+\frac{4}{S_\star T_\star^3}\ddot{\bar{x}}^0(\tau)\ddot{\bar{x}}^0(\tau')\,,
\end{equation} 
where the second term derives from fluctuations around $T_\star$~\eqref{eq:T_saddle}.

The operator $M$ has a zero mode because the solution spontaneously breaks the worldline time translation invariance, and for any solution $\bar{x}^0(\tau)$ there exists a translated solution $\bar{x}^0(\tau-\tau_0)$. We can see this explicitly from the fact that for an infinitesimal translation $\bar{x}^0(\tau-\tau_0)=\bar{x}^0(\tau)-\tau_0 \,\dot{\bar{x}}^0+\dots$ and noting that $\dot{\bar{x}}^0$ is a zero mode of $M$
\begin{equation}
\int_0^1 d\tau' M(\tau,\tau')\dot{\bar{x}}^0(\tau') = 0\,.
\end{equation}
For the diagonal term, this follows from the time derivative of our background equation of motion~\eqref{eq:insteom}. For the second term it follows from the fact that the integrand is a total derivative and the background solution is periodic.

Let us now decompose the fluctuations into eigenfunctions of $M(\tau,\tau')$ and isolate this zero mode. The first term in $M(\tau,\tau')$ is the same operator we considered in section~\ref{sec:slowmass}, but now acting on periodic functions. For periodic functions, the expansion of the fluctuations around the background $\bar{x}(\tau-\tau_0)$ involves two linearly independent sets of eigenmodes
\begin{equation}
\delta x^0(\tau) = \sum_{k=1}^\infty  a_k \delta x^{(1)}_k(\tau-\tau_0)+\sum_{k=0}^\infty b_k \delta x^{(2)}_k(\tau-\tau_0)\,.
\end{equation}
We will define the first set so that $\delta x^{(1)}_k(0)=0$ and the second set so $\delta x^{(2)\prime}_k(0)=0$, and we define the labels so that modes with label $k$ have $2k$ zeros. We normalize them so that 
\begin{equation}
\int_0^1 d\tau \,T_\star \delta x^{(i)}_k(\tau) \delta x^{(j)}_l(\tau) = \delta^{ij}\delta_{kl}\,.
\end{equation}
In terms of the expansion coefficients $a_k$ and $b_k$, the measure in $\zeta$-function regularization is given by
\begin{equation}
\mathcal{D}\delta x^0 = \prod_{k=1}^\infty \frac{da_k}{\sqrt{2\pi}}\prod_{k=0}^\infty\frac{db_k}{\sqrt{2\pi}}\,.
\end{equation}

For the solution that oscillates back and forth between the complex pair of zeros of the energy $n$ times, we see that the zero mode (like the background solution) has $2n$ zeros, and vanishes at $\tau = 0$ (for $\tau_0=0$). So the zero mode corresponds to $\delta x^{(1)}_n(\tau)$.  From the non-local action and the constraint, we find that the properly normalized mode is given by 
\begin{equation}
  \delta x^{(1)}_n(\tau) = \frac{i}{\sqrt{S_\star} T_\star}\dot{\bar{x}}^0(\tau)\,.\label{eq:zeromode}
\end{equation}
Changing variables from $a_n$ to $\tau_0$, the measure becomes
\begin{equation}
 \mathcal{D}\delta x^0 = i\frac{\sqrt{S_\star} T_\star}{\sqrt{2\pi }} d\tau_0 {\prod_{\substack{k=1\\ k\neq n}}^\infty} \frac{da_k}{\sqrt{2\pi}}\prod_{k=0}^\infty\frac{db_k}{\sqrt{2\pi}}\,,
\end{equation}
The integration range for $\tau_0$ must be chosen so that the phase of the solution varies over one period, i.e. $0\leq \tau_0\leq 1/n$, and because the action is independent of $\tau_0$, the integral simply yields a factor $1/n$. Analytically continuing the integration variables, recalling again that the original $T$ integration contour had a small positive imaginary part to determine the direction, the effective action can then be written as
\begin{equation}
  iW = \frac{V}{4}\int\frac{d^3p}{(2\pi)^3}\sum_\mathrm{saddles} \frac{1}{n}\exp\left(-S_\star[\bar{x}^0]\right)\sqrt{\frac{\det' M_0}{\det' M}} \,.
\end{equation}
Here we have introduced $M_0= -(1/T_\star^2)\, d^2/d\tau^2$ and have used that in $\zeta$-function regularization $\sqrt{\det' M_0}=T_\star$ to express the factor $T_\star$ from the Jacobian associated with the transformation from $a_n$ to $\tau_0$ in terms of the square root of the determinant. The primes as usual indicate that the zero mode has been omitted in the computation of both determinants. The ratio of functional determinants could again be evaluated in terms of a generalization of the method of Gelfand and Yaglom to periodic boundary conditions or in a heat-kernel expansion. Let us instead evaluate it by deriving the spectrum in the limit in which $\omega(\tau)$ varies slowly over one period, working in the WKB approximation. As we show in Appendix~\ref{app:ratio}, in this approximation the square root of the ratio of functional determinants simply evaluates to 
\begin{equation}
  \sqrt{\frac{\det' M_0}{\det' M}} = -(-1)^{n-1}.
\end{equation}
Even without a detailed calculation, note that the oscillation theorem and the fact that $\delta x_n^{(1)}(\tau)$ is a zero mode implies that the eigenvalues associated with the modes with $k<n$ are negative whereas those with $k>n$ are positive. This gives rise to a factor of $-i(-1)^{n-1}$. As we show in Appendix~\ref{app:ratio}, the second term in $M(\tau,\tau')$ ensures that the eigenvalue associated with the remaining mode, $\delta x^{(2)}_n(\tau)$, is also negative so that the ratio of determinants will contain a factor $-(-1)^{n-1}$. The calculation presented in the appendix shows that it equals $-(-1)^{n-1}$.

Using that the $n$-instanton action for the $i$-th production event is simply $n$ times the instanton action $S_\star[\bar{x}^0]=n S_{\rm inst, i}$, the imaginary part of the effective action takes the simple form
\begin{equation}
  \mathrm{Im} W = \frac{1}{4}\int d^3 x\int\frac{d^3p}{(2\pi)^3}\sum_i \ln\left(1+e^{-S_{\rm inst, i}}\right) \,,\label{eq:worldline_saddle_final}
\end{equation}
where the sum runs over all production events, and the instanton action for the $i$-th event is given by~\eqref{eq:inst_contour}. This agrees with and generalizes our earlier result~\eqref{eq:ImW_slowmass}.

\section{Two-point function in quantum field theory}\label{sec:propagator}
As we saw, the vacuum persistence amplitude can be used to obtain occupation numbers, but it does not provide information about the spatial distribution of particles when they are produced. This is phenomenologically interesting, and we will now extract it from the two-point correlation function. We will derive the two-particle wavefunction in quantum field theory in this section and then turn to the worldline description in section~\ref{sec:worldline_prop}. 

The two-particle wavefunction $\Psi(\mathbf{x}_1,\mathbf{x}_2)$ as measured by a late-time observer assuming that the system was in the vacuum of the early-time observer is contained in the two-point correlation function
\begin{equation}
\Psi(\mathbf{x}_1,\mathbf{x}_2) \subset\mathcal{N} \lim_{t_1 \rightarrow t_2}\langle \chi(x_1)\chi(x_2)\rangle \,, \label{2ptfnc}
\end{equation}
where $\mathcal{N}$ is a normalization constant, and we have defined
\begin{equation}
  \langle \chi(x_1)\chi(x_2)\rangle =\frac{\langle 0,\mathrm{out}| \chi(x_1)\chi(x_2)|0, \mathrm{in}\rangle}{\langle 0,\mathrm{out} |0, \mathrm{in}\rangle} \,.
\end{equation} 
As before, we write the $|0,\mathrm{in}\rangle$ state as a superposition of excited $out$-particle states
\begin{multline}
  \langle \chi(x_1)\chi(x_2)\rangle =\\ 
\langle 0,\mathrm{out} |\chi(x_1)\chi(x_2)|0, \mathrm{out}\rangle + \int\frac{d^3 p}{(2\pi)^3}\, \frac{\beta_p}{2\alpha_p^*} \langle 0, \mathrm{out}|\chi(x_1)\chi(x_2) a^\dagger_{\mathbf{p}} a^\dagger_{-\mathbf{p}} |0,\mathrm{out}\rangle + \dots\,. \label{inoutexpansion}
\end{multline}
The two terms in the expansion given here are the two-point vacuum correlation function, and the amplitude for the process in which two $out$-particles are subsequently annihilated by the field at $\mathbf{x}_1$ and $\mathbf{x}_2$. The wavefunction does not include the vacuum contribution, but we will keep track of it for now. The terms in the sum represented by dots correspond to multiple pairs that would contribute if the $\chi$-particles were interacting. 

Expanding the fields in terms of their Fourier modes, as in \eqref{eq:chi_fourier}, we have
\begin{equation}
  \langle \chi(x_1)\chi(x_2)\rangle = \int \frac{d^3p}{(2\pi)^3} e^{i\mathbf{p}\cdot(\mathbf{x}_1 - \mathbf{x}_2)} \,\left(|u^{\rm out}_p(t)|^2 + \frac{\beta_p}{\alpha_p^*} u^{\rm out}_{p}(t)^2 \right)\,, \label{wavefunctionpropagator}
\end{equation} 
and we see that the two-particle wavefunction corresponds to the positive frequency part of the two-point correlation function, and its Fourier transform is simply given by 
\begin{equation}
\Psi(\mathbf{p},t)=\mathcal{N} \frac{\beta_p}{\alpha_p^*} u^{\rm out}_p(t)^2 \,.
\end{equation}
This will be useful for comparison with the computation in first quantization in section~\ref{sec:worldline_prop}. 

We are most interested in the spatial separation of the quanta at the time of production. Particle production occurs on a timescale set by the distance between the maxima of the dimensionless parameter $|\dot{\omega}_p/\omega_p^2|$, where $\omega_p^2 = p^2 + m(t)^2 $. However, it is sufficient for our purpose to compute the wavefunction at the instant the mass minimized. (See Appendix~\ref{app:time} for more details.)
\begin{align}
  \Psi(\mathbf{x}_1,\mathbf{x}_2)=\mathcal{N}\int \frac{d^3p}{(2\pi)^3} e^{i\mathbf{p}\cdot(\mathbf{x}_1 - \mathbf{x}_2)} \frac{\beta_p}{\alpha_p^*} u^{\rm out}_p(0)^2 \,.
\end{align}

So far, this was general. For the model with quadratic time dependence~\eqref{eq:moft}, the mode functions are parabolic cylinder functions, normalized to ensure canonical commutation relations. In particular,
\begin{equation}
u_{p}^{\rm out}(t) = \frac{e^{-\pi E_p^2/8\omega}}{(2\omega)^{1/4}}\;D_{-\frac{1}{2} -\frac{iE_p^2}{2\omega}}(e^{i\pi/4}\sqrt{2\omega}t)\,, \label{uoutmodefnc}
\end{equation}
where $ E_p = \sqrt{m_0^2 + p^2}$ and $\omega=g\dot\phi\approx$ const.

Using $D_\nu(0)=2^{\nu/2}\sqrt{\pi}/\Gamma((1-\nu)/2)$ and the Bogoliubov coefficients derived in Appendix~\ref{app:exact}, the two-particle wavefunction can be simplified to
\begin{equation} 
\Psi(\mathbf{x}_1,\mathbf{x}_2)=\frac{(-1)^{3/4}\mathcal{N}}{8\pi^2 \sqrt{\omega}} \int_0^\infty p dp \, \frac{\sin pr}{r} \, e^{-\frac{\pi E_p^2}{2\omega}}\, \frac{\Gamma(\frac{1}{4}+\frac{i E_p^2}{4\omega})}{\Gamma(\frac{3}{4}+\frac{i E_p^2}{4\omega})}\,, \label{qft_wf}
\end{equation}
where $r=|\mathbf{x}_1-\mathbf{x}_2|$. We see that $|\Psi(r)|$ regular as $r\to 0$. As $r \to \infty$, the wavefunction decays exponentially. The detailed behavior of the tail depends on the value of the instanton action.

For real $p$ and large instanton action, the magnitude of the argument of the $\Gamma$-functions is large, and their ratio is well approximated by
\begin{equation}
\frac{\Gamma(\frac{1}{4}+\frac{i E_p^2}{4\omega})}{\Gamma(\frac{3}{4}+\frac{i E_p^2}{4\omega})} \approx e^{-i\frac{\pi}{4}}\, \frac{2\sqrt{\omega}}{ E_p}\label{eq:gammaapprox}\,.
\end{equation}
For $\sqrt{\omega}r\lesssim\sqrt{\pi}\sqrt{S_{\rm inst}}$, we can approximate $ E_p\approx m_0$  in the denominator, and the resulting integral becomes
\begin{equation}
\Psi(r)= -\frac{ \sqrt{2}\mathcal{N}}{(2\pi)^3}\frac{\omega^{3/2}\, e^{-\frac{m_0^2 \pi}{2\omega}}}{m_0}\, e^{-\frac{\omega r^2}{2\pi}}\label{eq:qft_wf1}
\end{equation}
For $\sqrt{\omega}r\gtrsim\sqrt{\pi}\sqrt{S_{\rm inst}}$, the location of the saddle point is dominated by the presence of $E_p$ in the denominator in~\eqref{eq:gammaapprox} and the wavefunction approaches
\begin{equation}
\Psi(r)= -\frac{\mathcal{N}}{(4\pi)^{3/2}}\frac{\sqrt{m_0}}{r^{3/2}}\, e^{-m_0 r}\,.
\end{equation} 

When the instanton action is large, the wavefunction around the transition is suppressed by $e^{-S_{\rm inst}/2}$ relative to its value at the peak. So when used to calculate moments, for large instanton action the wavefunction is well approximated by a Gaussian over the full range of interest  
\begin{equation}
\Psi(r) \propto \exp\left(-\frac{\omega r^2}{2\pi} \right)\label{eq:gaussianwavefunction}\,.
\end{equation}

We can now extract the average separation $\langle r \rangle$ between the produced particles around the time of production, the most likely separation $\hat{r}$, as well as the width $\sigma_r$, etc. For the mean we find
\begin{equation}
\langle r\rangle \approx \frac{\int_0^\infty dr\, r^3 \exp\left(-\frac{\omega r^2}{\pi}\right)}{\int_0^\infty dr\, r^2 \exp\left(-\frac{\omega r^2}{\pi}\right)} = \frac{2}{\sqrt{\omega}}\,,
\end{equation}
and the expected scatter around the mean is
\begin{equation}
\sigma_r = \sqrt{\langle r^2 \rangle - \langle r\rangle^2} \approx \frac{1}{\sqrt{\omega}}\,.
\end{equation}
The most likely separation is given by the location of peak of $r^2|\Psi(r)|^2$
\begin{equation}
\hat{r} = \sqrt{\frac{\pi}{\omega}}\,.~\label{rhat particle}
\end{equation}

\section{Two-point function in first quantization} \label{sec:worldline_prop}
We will now derive the propagator and two-particle wavefunction in first quantization. Proceeding as we did for the effective action, we can write the Green's function as 
\begin{equation}
G(x_f,x_i) = \frac12\int_0^\infty dT\, \langle x_f^\mu|\exp\left(-T H\right)|x_i^\mu\rangle\,,\label{eq:propagator}
\end{equation}
where the Hamiltonian is given as before by~\eqref{eq:hamiltonian}.
We will consider this Hamiltonian formulation in subsection~\ref{ssec:hamiltonian2pt}. This will give us another way to derive the occupation numbers, including for masses with a time-dependence for which our earlier approximations are insufficient. 

We can equivalently write the Green's function as a worldline path integral 
\begin{equation}
G(x_f,x_i) = \frac12\int_0^\infty dT \int_{ x^\mu_i}^{ x^\mu_f} \mathcal{D} x^\mu \exp\left[{-\int_0^1 d\tau \left(\frac{1}{2T}\dot{x}_\mu\dot{x}^\mu +\frac{T}{2}m^2(x)\right)}\right]\,,
\end{equation}
summing over all paths that start at $x_i^\mu$ at $\tau=0$ and end at $x^\mu_f$ at $\tau=1$, and we will consider this in subsection~\ref{ssec:worldline2pt} to derive the two-particle wavefunction.

\subsection{Two-point correlation function and Bogoliubov coefficients}\label{ssec:hamiltonian2pt}
If we assume that the mass only depends on a single coordinate, in our case the time coordinate, we can insert a complete set of momentum states into equation~\eqref{eq:propagator} to write the Green's function as\footnote{The small imaginary part in the denominators in~\eqref{eq:energy_decomposition} can most easily be obtained by working with Minkowski time and then analytically continuing to Euclidean time.}
\begin{equation}
  G(x_f,x_i) = \frac12\int_0^\infty dT \int \frac{d^3 p}{(2\pi)^3} e^{i\mathbf{p}\cdot(\mathbf{x}_f-\mathbf{x}_i)}\langle x_f^0|\exp\left(T H_p\right)|x_i^0\rangle\,,\label{eq:propagator_p}
\end{equation}
where the Hamiltonian is given by
\begin{equation}
  H_p = \frac12 p_0^2-\frac12 E_p(x^0)^2\,.
\end{equation}
We see that this describes a particle in the potential $-E_p(x^0)^2/2$. We can gain insight into the behavior of the Green's function by inserting a complete set of energy eigenstates. Assuming that the energy is positive definite as $|x^0|\to\infty$, these will consist of a continuum as well as possibly bound states. If the mass asymptotes to $m^2(x^0)\to m^2_\pm$ as $x^0\to\pm\infty$,\footnote{We allow $m^2_\pm$ to be infinite as would be the case for the quadratic time dependence.} there will be a single energy eigenstate for energies between $-m^2_+/2$ and $-m^2_-/2$ and a doubly degenerate continuum of states for energies above this range. We will denote the continuum states by $|\mathcal{E}, \sigma \rangle$, where $\sigma$ labels the two energy eigenstates, and the bound state with energy $\mathcal{E}_n$ by $|n \rangle$, and we assume they are normalized so that
\begin{equation}
  \langle \mathcal{E}, \sigma |\mathcal{E}', \sigma' \rangle = \delta(\mathcal{E}-\mathcal{E}')\delta_{\sigma\sigma'}\,,\quad \langle n |{n'} \rangle = \delta_{nn'}\,.\label{eq:normalization}
\end{equation}
After performing the integral over the Schwinger parameter, the Green's function can be written as
\begin{eqnarray}
  G(x_f,x_i) &=& -\frac{1}{2}\int \frac{d^3 p}{(2\pi)^3} e^{i\mathbf{p}\cdot(\mathbf{x}_f-\mathbf{x}_i)}\nonumber\\
  &&\times\left[\sum_{\sigma} \int_{-E_\sigma^2/2}^{\infty}d\mathcal{E} \frac{ \langle x_f^0|\mathcal{E},\sigma\rangle \langle \mathcal{E},\sigma|x_i^0\rangle}{\mathcal{E}+i\epsilon}+\sum_{n} \frac{\langle x_f^0|n\rangle\langle n|x_i^0\rangle}{\mathcal{E}_n+i\epsilon} \right] \,, \label{eq:energy_decomposition}
\end{eqnarray}
where the lower limits on the integrals are given by $E_\pm^2=p^2+m_\pm^2$. As discussed earlier, if $m_+^2$  and $m_-^2$  differ, there will be a range of energies for which there is only a single state. 

The bound state wavefunctions decay exponentially away from the production region, and we see that while they are needed to understand the Green's function and the wavefunction around the time of production, the continuum states are sufficient to understand the Green's function and wavefunction long before and after the production occurs.\footnote{The bound states would also play a role if we were interested in the behavior of the Green's function for complex times.}

The continuum wavefunctions obey the Schr\"odinger equation
\begin{equation}
  \left(-\frac12\frac{d^2}{dx^0} - \frac12 E^2_p(x^0)\right)\langle x^0|\mathcal{E},\sigma\rangle = \mathcal{E}\langle x^0|\mathcal{E},\sigma\rangle\,,\label{eq:contwavefn}
\end{equation}

Provided $E_p^2(x^0)$ is positive long before and after the production event, at early and late times the energy eigenfunctions will be well described by the WKB approximation. Let us assume that the solution with $\sigma=+$ is the positive frequency solution at early times so that as $x^0\to-\infty$ 
\begin{equation}
  \langle x^0|\mathcal{E},+\rangle = \frac{t(\mathcal{E})}{\sqrt{2\pi}\left(2\mathcal{E}+E^2_p(x^0)\right)^{1/4}} \exp\left(-i\int^{x^0}_{x^0_1} d x \sqrt{2\mathcal{E}+E^2_p(x)}\right)\,, 
\end{equation} 
where for later convenience, we have chosen the time of the first production event as the lower limit of integration. The Stokes phenomenon implies that at late times the solution will acquire a small negative frequency part
\begin{eqnarray}
  \langle x^0|\mathcal{E},+\rangle &=& \quad\frac{1}{\sqrt{2\pi}\left(2\mathcal{E}+E^2_p(x^0)\right)^{1/4}} \exp\left(-i\int^{x^0}_{x^0_1} d x \sqrt{2\mathcal{E}+E^2_p(x)}\right) \nonumber\\
  &&+\,\frac{r(\mathcal{E})}{\sqrt{2\pi}\left(2\mathcal{E}+E^2_p(x^0)\right)^{1/4}} \exp\left(+i\int^{x^0}_{x^0_1} d x \sqrt{2\mathcal{E}+E^2_p(x)}\right)\,,
\end{eqnarray} 
where we have used the notation familiar from quantum mechanics for the transmission amplitude, $t$, needed to ensure the wavefunction is normalized according to~\eqref{eq:normalization}, and for the reflection amplitude $r$, which satisfy $|t|^2+|r|^2=1$. We can think of this as a scattering problem in which a beam of particles that is incoming from plus infinity is partially reflected and partially transmitted. The orthogonal solution with $\sigma=-$ corresponds to the scattering problem in which the beam is incident from minus infinity. 

Let us now specialize to the two-point correlation function at late times, setting $x_f^0=x_i^0=x^0$. Performing the integral, for example, by changing the integration variable from $\mathcal{E}$ to $p_0=\sqrt{2\mathcal{E}+E^2_p(x^0)}$, and deforming the contour, we see that, up to terms that vanish exponentially for large $x^0$, the equal time two-point correlation function at late times becomes
\begin{eqnarray}
  G(\x_f,\x_i,x^0) &=& i\int \frac{d^3 p}{(2\pi)^3} e^{i\mathbf{p}\cdot(\mathbf{x}_f-\mathbf{x}_i)}\nonumber\\
  &&\times\left[\frac{1}{2 E_p(x^0)}+\frac{r^*(0)}{2E_p(x^0)} \exp\left(-2i\int^{x^0}_{x^0_1} d x\, E_p(x)\right) \right] \,. \label{eq:propagator2}
\end{eqnarray}
Comparing with the quantum field theory result~\eqref{wavefunctionpropagator}, we see that the $out$-mode function is given by 
\begin{equation}
  u^{\rm out}_p(x^0) = \frac{1}{\sqrt{2E_p(x^0)}} \exp\left(-i\int^{x^0}_{x^0_1} d x\, E_p(x)\right)\,,
\end{equation}
which agrees with the WKB solution of equation~\eqref{eq1}. More importantly, the occupation number for the mode with momentum $\mathbf{p}$ is determined by the reflection coefficient $R=|r(0)|^2$ for scattering off the potential $-E^2_p(x^0)/2$ from plus infinity, evaluated at energy $\mathcal{E}=0$  
\begin{equation}
  |\beta_p|^2 = \frac{R}{1-R}\,.
\end{equation}
This provides another way to compute the occupation numbers~\cite{Audretsch:1979uv}, including in regimes where our earlier approximations may be invalid. In particular, it means that we can immediately obtain the occupation numbers for any functional form of the mass for which the reflection coefficient for the potential $-E^2_p(x^0)/2$ is known. It also means we can employ well-known approximation techniques like the Born approximation or eikonal approximations used in scattering theory to compute the occupation numbers if the exact form is not known.

For positive definite $E_p(x^0)$, scattering with $\mathcal{E}=0$ amounts to the quantum mechanical reflection for scattering above a barrier discussed in the textbook by Landau and Lifshitz~\cite{Landau1981Quantum}. Comparing with our earlier results, we see that for a single production event at $x_i^0$ and small occupation numbers
\begin{equation}
  R = e^{-S_{\rm inst}}
\end{equation}
and the instanton action is given by 
\begin{equation}
  S_{\rm inst} =2\int_{-x_i^4}^{x_i^4} dx^4\, E_p(x_i^0+i x^4)\,,
\end{equation}
rederiving the reflection coefficient $R$ given there. The fact that the coefficient of the exponential is unity in our derivation follows from the result derived in Appendix~\ref{app:ratio} that the magnitude of the ratio of functional determinants is unity. 

In the limit of small occupation numbers, we saw that for multiple production events the occupation number of the individual events simply add up. 

These results also allow us to give a general expression for the two-particle wavefunction. Up to a phase, $r(0)^* = e^{-S_{\rm inst}/2}$, so that according to~\eqref{eq:propagator2} the two-particle wavefunction sufficiently long after the first production event (but before the second production event) is simply given by
\begin{equation}
  \Psi(\mathbf{x}_1,\mathbf{x}_2, x^0) = \mathcal{N}\int \frac{d^3 p}{(2\pi)^3} e^{i\mathbf{p}\cdot(\mathbf{x}_1-\mathbf{x}_2)}\frac{e^{-S_{\rm inst}/2}}{2E_p(x^0)} \exp\left(-2i\int^{x^0}_{x^0_1} d x\, E_p(x)\right) \,. 
\end{equation}
The exponential of the instanton action is associated with the production event, and the phase factor describes the spreading of the wavefunction after the production event associated with the free propagation of the particles. 

As an approximation to the wavefunction at the time of production, we can undo the free propagation by stripping off the phase factor and take 
\begin{equation}
  \Psi(\mathbf{x}_1,\mathbf{x}_2, x^0_1) \approx \mathcal{N}\int \frac{d^3 p}{(2\pi)^3} e^{i\mathbf{p}\cdot(\mathbf{x}_1-\mathbf{x}_2)}\frac{e^{-S_{\rm inst}/2}}{2E_p(x^0_1)}\,.\label{eq:wavefcn_approx}
\end{equation}
This is exact if the WKB approximation holds throughout, but will in general receive corrections. 

\subsection{Two-particle wavefunction from the worldline path integral}\label{ssec:worldline2pt}

Let us now turn to the path integral representation and derive the two-particle wavefunction. Starting from equation~\eqref{eq:propagator_p}, we express the Green's function as a path integral 
\begin{multline}
  G(\x_f,\x_i,t) = \int\frac{d^3 p}{(2\pi)^3} e^{i\mathbf{p}\cdot(\mathbf{x}_f-\mathbf{x}_i)}\\\times\frac12\int_0^\infty dT \int_{t}^{t} \mathcal{D} x^0 \exp\left[-\int_0^1 d\tau \, \left(-\frac{1}{2T}(\dot{x}^0)^2 + \frac{T}{2}E_p(x^0)^2\right) \right]\,.
\end{multline}
Our goal here will not be to rederive the full Green's function, only to extract the approximation of the wavefunction~\eqref{eq:wavefcn_approx} derived in the last subsection. For this purpose, let us integrate over the momenta associated with the spatial directions  
\begin{multline}
  G(\x_f,\x_i,t) =\\ \frac12\int_0^\infty \frac{dT}{(2\pi T)^{3/2}} \int_t^t \mathcal{D} x^0 \exp\left[-\int_0^1 d\tau \, \left(-\frac{1}{2T}(\dot{x}^0)^2 +\frac{1}{2T} r^2+ \frac{T}{2}E_p(x^0)^2\right) \right]\,,
\end{multline}
where $r = |\x_f-\x_i|$. Integrating over the Schwinger parameter in the saddle point approximation, this becomes
\begin{equation}
  G(\x_f,\x_i,t) = \frac{1}{4\pi}\sum_{\rm saddles} \int_t^t \mathcal{D} x^0  \frac{\exp\left(-S_\star[x^0]\right)}{\sqrt{S_\star[x^0]T_\star[x^0]}}\,,
\end{equation}
where the non-local action is now given by 
\begin{equation}
  S_\star[x^0] = \sqrt{\int_0^1\,d\tau\,\left(r^2-(\dot{x}^0)^2\right)} \,\sqrt{\int_0^1\,d\tau\, m^2}\,,
\end{equation}
and $T_\star$ is given by
\begin{equation}
T_\star[x] = \frac{\sqrt{\int\,r^2-(\dot{x}^0)^2}}{\sqrt{\int m^2}}\,.\label{eq:taustarint}
\end{equation}
In the limit in which the non-local action is large, we can also perform the integral over the coordinate $x^0$ in a saddle point approximation. The variation of the action with respect to $x^0$ leads to the equation of motion
\begin{equation}
\frac{\delta S_\star}{\delta x^0}  = \frac{1}{T_\star[x]}\frac{d^2x^0}{d\tau^2} + \frac12\frac{dm^2}{dx^0} T_\star[x] = 0\,,\label{eq:propeom}
\end{equation} 
and for the equal-time correlator, we are looking for a solution with boundary conditions $\bar{x}^0(0) = \bar{x}^0(1)=t$. 

Decomposing the path into the background and fluctuations around it, the equal-time two-point correlation function can be written as 
\begin{multline}
G(\x_i,\x_f,t) = \frac{1}{4\pi}\sum_{\rm saddles}\frac{e^{-S_\star[\bar{x}]}}{\sqrt{S_\star[\bar{x}] T_\star[\bar{x}]}} \\\times\int{\mathcal{D}\delta x^0} \exp\left[\frac12\int d\tau d\tau' \delta x^0(\tau)  M(\tau,\tau')\delta x^0(\tau')+\dots\right]\,,
\end{multline}
where again $M$ is the second functional derivative of the non-local action $S_\star$ with respect to $x^0$ evaluated on the background solution $\bar{x}^0(\tau)$
\begin{equation}
M=\left(-\frac{1}{T_\star}\frac{d^2}{d\tau^2}-\frac{T_\star}{2}\left.\frac{d^2m^2}{{dx^0}^2}\right|_{\bar{x}}\right)\delta(\tau-\tau')+\frac{4}{S_{\star}[\bar{x}]T_\star^2}\ddot{\bar{x}}^0(\tau)\ddot{\bar{x}}^0(\tau')\,.
\end{equation}

If the mass is time-independent, there is only one solution for equation~\eqref{eq:propeom} with the required boundary conditions, $\bar{x}^0(\tau)=t$. For this solution $S_\star[\bar{x}]=mr$ and $T_\star[\bar{x}]=r/m$. Performing the integral over the fluctuations around this solution yields the familiar behavior of the two-point vacuum correlation function at spacelike separation $G(x_i,x_f)\propto m^{1/2}\exp(-mr)/r^{3/2}$. 

If the mass has a minimum, and we are interested in the correlation function at the time $t_0$ at which the mass is minimized, there is again a solution to the equation of motion~\eqref{eq:propeom}, $\bar{x}^0(\tau)=t_0$.  After integrating over the fluctuations, the contribution from the constant solution yields the vacuum two-point correlation function. In addition, there are now also non-trivial solutions that yield the two-particle wavefunction. 

For the model with quadratic time-dependence, the background solutions are again trigonometric functions, and given our boundary conditions, we now have solutions with $T_\star=n\pi /b$ so that\footnote{Notice that the boundary conditions have eliminated the zero mode associated with $\tau_0$.}
\begin{equation}
\bar{x}^0(\tau) = \sqrt{\left(\frac{r}{n\pi}\right)^2-\left(\frac{a}{b}\right)^2}\sin(n\pi\tau)\,,
\end{equation}
and the non-local action evaluated on these solutions is 
\begin{equation}
S_\star[\bar{x}]=\frac{n\pi a^2}{2b}+\frac{br^2}{2\pi n}\,.
\end{equation}
For $\sqrt{b}r\lesssim \sqrt{\pi}\sqrt{S_{\rm inst}}$, the leading saddle corresponds to $n=1$, and the leading exponential behavior agrees with our earlier calculation~\eqref{eq:gaussianwavefunction}, $\Psi(r)\propto \exp(-b r^2/2\pi)$.  For larger values of $r$, saddles with $n\sim \sqrt{b}r/\sqrt{S_{\rm inst}}\sqrt{\pi}$ will dominate, and the wavefunction in the tail decays like a simple exponential $\Psi(r)\propto \exp(-a r)$, also reproducing our earlier result.

To obtain the prefactor, we perform the integral over the fluctuations. For the expansion
\begin{equation}
\delta x^0(\tau) = \sum_{m=1}^\infty a_m\sqrt{\frac{2}{T_\star}}  \sin(m\pi\tau)\,,
\end{equation}
the measure in $\zeta$-function regularization is given by 
\begin{equation}
\mathcal{D}\delta x^0 = \frac{1}{\sqrt{\pi}}\prod_{m=1}^\infty \frac{da_m}{\sqrt{2\pi}}\,.
\end{equation}
In the limit of large instanton action and for $\sqrt{b}r\lesssim \sqrt{\pi}\sqrt{S_{\rm inst}}$ where the leading saddle dominates, $S_\star=\pi a^2/2b$ and $T_\star=\pi/b$. Accounting for the phase in the definition of the Green's function that can be obtained from the theory of a free massive scalar to be $-e^{i\pi/4}$, we have
\begin{equation}
G(x_i,x_f)\approx i\frac{\sqrt{2}}{(2\pi)^3}\frac{b^{3/2}}{a}e^{-\frac{\pi a^2}{2b}}e^{-\frac{b r^2}{2\pi}}\,.
\end{equation} 
Making use of the relation $\langle\chi(x)\chi(y)\rangle=iG(x,y)$, this reproduces the two-particle wavefunction~\eqref{eq:qft_wf1}.

Of course, for the model with quadratic time-dependence, we can perform the path integral over the fields exactly because the action is quadratic, as we did in Appendix~\ref{app:propagator}. We have included this way to compute the wavefunction from the path integral because we cannot in general evaluate the path integral exactly. 

\section{Future directions}
In this paper, we revisited the problem of particle production in a cosmological setting.  We focused on the vacuum persistence amplitude and the propagator that contain the desired information about the produced particles, in particular occupation numbers and the two-particle wavefunction. Both the vacuum persistence amplitude and the propagator can readily be computed in a first quantized formalism as Euclidean path integrals 
\begin{equation}
  \frac{1}{{\rm Vol}_{\rm Diff}}\int \mathcal{D}e\mathcal{D}x \,e^{-S[X,e]}\,,
\end{equation}
defined on a circle for the vacuum persistence amplitude and on an interval for the propagator. In this paper, we focused on the production of spin-0 particles, but the techniques can readily be extended to the production of spin-1/2 or spin-1 particles by including the appropriate degrees of freedom in the worldline action. 

Similarly, the techniques can be used to compute the production of extended objects like strings or branes. The equivalent Euclidean path integrals are given by 
\begin{equation}
  \frac{1}{{\rm Vol}_{\rm Diff}}\int \mathcal{D}g\mathcal{D}x \,e^{-S[X,g]}\,,
\end{equation}
and for fundamental strings we should further quotient by the volume of the group of Weyl transformations. We will discuss the production of extended objects in a future publication. 
 
In this paper, we focused on scenarios in which production events are well-separated in time and the corresponding instantons are free. It is natural to consider scenarios in which interactions between instantons become important.

\section*{Acknowledgments}
The authors are supported in part by the Department of Energy under Grant No. DE-SC0009919.

\appendix

\section{Time-dependent mass from expansion}
\label{app:mass_expansion}
In this appendix we show that the production of scalar particles in an expanding universe can be related to the production of particles with a time-dependent mass in Minkowski space. 

To see this, let us consider the partition function 
\begin{equation}
Z = \int \mathcal{D}_g\phi \exp\left[i\int d^4x\sqrt{-g} \left(-\frac12g^{\mu\nu}\partial_\mu\phi\partial_\nu\phi - \frac12 m^2\phi^2-\frac12\xi R\phi^2\right)\right]\,,
\end{equation}
where $R$ is the Ricci scalar, $\xi=0$ for a minimally coupled scalar field, and we define the measure so that 
\begin{equation}
\int \mathcal{D}_g\phi \exp\left[\frac{i}{2}\int d^4x \sqrt{-g} \phi^2\right]=1\,.
\end{equation}
Let us now consider an FLRW cosmology in conformal time $\tau$, so that the metric given by
\begin{equation}
g_{\mu\nu} = a^2(\tau)\eta_{\mu\nu}\,.
\end{equation}
We will assume that at early and late times there is a natural notion of a vacuum state. The partition function then computes the vacuum persistence amplitude 
\begin{equation}
  \langle 0,\mathrm{out}|0,\mathrm{in}\rangle = \int \mathcal{D}_a\phi \exp\left[i\int d^4x \, a^2(\tau) \left(-\frac12\eta^{\mu\nu}\partial_\mu\phi\partial_\nu\phi - \frac12 a^2(m^2+\xi R)\phi^2\right)\right]\,,
\end{equation} 
and the definition of the measure becomes 
\begin{equation}
\int \mathcal{D}_a\phi \exp\left[\frac{i}{2}\int d^4x\, a^4(\tau) \phi^2\right]=1\,.\label{eq:measure}
\end{equation}
The kinetic term of the scalar field action still differs from the action for a scalar field in Minkowski space by factor of $a^2(\tau)$. We can absorb it by performing a field redefinition $\phi=\varphi/ a(\tau)$. Dropping a total derivative, the vacuum persistence amplitude then becomes
\begin{equation}
\langle 0,\mathrm{out}|0,\mathrm{in}\rangle = \int \mathcal{D}\varphi\, J \exp\left[i\int d^4x \, \left(-\frac12\eta^{\mu\nu}\partial_\mu\varphi\partial_\nu\varphi - \frac12 a^2\left(m^2+\left(\xi-\frac16 \right)R\right)\varphi^2\right)\right]\,, 
\end{equation}
where $J$ is the Jacobian associated with the field redefinition, and we define the measure for the field $\varphi$ as we would in Minkowski space 
\begin{equation}
\int \mathcal{D}\varphi \exp\left[\frac{i}{2}\int d^4x \,\varphi^2\right]=1\,.
\end{equation} 
Substituting the field redefinition and change of variables into equation~\eqref{eq:measure}, we find that the Jacobian is given by 
\begin{equation}
J = \sqrt{\det\left(a^2(\tau)\delta^{4}(x-x')\right)}=\exp\left[\frac12\delta^4(0)\int d^4 x \ln a^2(\tau)\right]\,,
\end{equation}
where the $\delta$-function evaluated at the origin appears because the transformation was local. This makes it convenient to use a regulator for which $\delta(0)=0$ so that the Jacobian is unity, $J=1$. In our case, we will work with a combination of $\zeta$-function regularization to evaluate mode sums associated with the time-direction, and dimensional regularization for the spatial components (of target space). The vacuum persistence amplitude for a field in an FLRW spacetime then simply becomes the vacuum persistence amplitude for a scalar field in Minkowski space with a time-dependent mass

\begin{equation}
  \langle 0,\mathrm{out}|0,\mathrm{in}\rangle = \int \mathcal{D}\varphi\, \exp\left[i\int d^4x \, \left(-\frac12\partial_\mu\varphi\partial^\mu\varphi - \frac12 m^2_{\rm eff}(\tau)\varphi^2\right)\right]\,, 
  \end{equation}
where the effective mass is given by
\begin{equation}
  m^2_{\rm eff}(\tau) = a^2(\tau)\left(m^2+\left(\xi-\frac16 \right)R\right)\,.
\end{equation}
As a simple consistency check, note that no particle production occurs for massless conformally coupled scalar fields in an expanding universe, $m=0$ and $\xi=1/6$, as expected.

\section{Exact solution for mass with quadratic time dependence}\label{app:exact}
Consider mode functions that satisfy the equation
\begin{equation}
\ddot{u}_{p}(t) + (E_p^2 + \omega^2 t^2)u_{p}(t)=0\,,\label{eq:mode_eq}
\end{equation}
where $E_p^2=p^2+m^2$ and $\omega$ is a constant. 
The general solution can be written in terms of the parabolic cylinder functions $D_\nu(z)$ as
\begin{equation}
u_{p}(t) = c_1 D_{-\bar\nu-1}(iyt) + c_2 D_{\bar\nu} (-yt)
\end{equation}
where
\begin{equation}
\quad y = e^{-i\frac{\pi}{4}}\sqrt{2\omega},\quad \bar\nu = -\frac{1}{2} + i\frac{E_p^2}{2\omega}
\end{equation}
Setting $c_1=c_2=e^{-\pi E_p^2 / 8\omega}(2\omega)^{-1/4}$, the Wronskian gives $W[u_p(t),u_p^*(t)] = i$ for each mode. Using the asymptotic form $D_\nu(z)\sim e^{-z^2/4}z^\nu$, valid provided $|\arg(z)|<3\pi/4$, we can identify the positive frequency solutions in the limits $t\to\pm\infty$. The $in$ and $out$ solutions with their WKB form are
\begin{align}
u_{p}^{\rm in}(t) &= \frac{e^{-\pi E_p^2/8\omega}}{(2\omega)^{1/4}}\;D_{\bar\nu}(-y t) \qquad \stackrel{t\to -\infty}{\sim} \quad \frac{e^{+\frac{1}{2}i\omega t^2}}{\sqrt{2\omega t}}\,,\\
u_{p}^{\rm out}(t) &= \frac{e^{-\pi E_p^2/8\omega}}{(2\omega)^{1/4}}\;D_{-{\bar\nu}-1}(i y t) \quad \stackrel{t\to +\infty}{\sim} \quad \frac{e^{-\frac{1}{2}i\omega t^2}}{\sqrt{2\omega t}} \,.\label{WKB_limits}
\end{align}
From equation~\eqref{eq:mode_eq}, we see that $D_\nu(-z)$ is a solution if $D_\nu(z)$ is. Since there are only two linearly independent solutions, there must then be a linear relation between $D_\nu(z)$, $D_\nu(-z)$, and $D_{-\nu-1}(iz)$. It is given by (see for instance \cite{GR}, equation 9.248)
\begin{equation}
D_\nu(-z) =  \frac{\sqrt{2 \pi}i}{\Gamma(-\nu)} e^{\frac{i\pi\nu}{2}} D_{-\nu-1}(i z)+ e^{i\pi \nu} D_\nu(z)\,.
\end{equation}
Under complex conjugation the parabolic cylinder functions satisfy $D_\nu(z)^* = D_{\nu^*}(z^*)$. So we see that $D_{-\bar\nu-1}(iyt)^*=D_{\bar\nu}(y t)$ and we have
\begin{equation}
u_p^{\rm in} = \frac{\sqrt{2 \pi}i}{\Gamma\left(-\bar\nu\right)}e^{\frac{i\pi\bar\nu}{2}} u_p^{\rm out} +e^{i\pi\bar\nu} u_p^{\rm out\, *}\,.
\end{equation}
Comparing with our definition of the Bogoliubov coefficients in~\eqref{eq:in_out}, we see that
\begin{equation}
\alpha_p = \frac{\sqrt{2 \pi}i}{\Gamma\left(-\bar\nu\right)}e^{\frac{i\pi\bar\nu}{2}}\qquad\text{and}\qquad \beta_p = -e^{i\pi\bar\nu}\,,
\end{equation}
or explicitly in terms of $E_p$ and $\omega$
\begin{equation}
\alpha_p = \frac{\sqrt{2 \pi i}\,e^{\frac{-\pi E_p^2}{4 \omega}}}{\Gamma\left(\frac{1}{2}-i\frac{ E_p^2}{2\omega}\right)}\qquad\text{and}\qquad \beta_p = i e^{\frac{-\pi E_p^2}{2 \omega}}\,.
\end{equation}

\section{Green's function for quadratic time dependence}\label{app:propagator}
Let us now consider the Green's function 
\begin{equation}
  G(\x_f,t_f,\x_i,t_i) = \frac12\int_0^\infty dT \int \frac{d^3 p}{(2\pi)^3} e^{i\mathbf{p}\cdot(\mathbf{x}_f-\mathbf{x}_i)}\langle t_f|\exp\left(T H_p\right)|t_i\rangle\,,
\end{equation}
for a mass with quadratic time dependence for which 
\begin{equation}
  H_p = \frac12 p_0^2-\frac12 E_p(t)^2\qquad\text{with}\qquad E_p(t)^2 = p^2+m^2+\omega^2 t^2\,.
\end{equation}

This example is instructive because all methods can be evaluated exactly.

\subsection{Hamiltonian formulation}
Let us first consider the Hamiltonian formulation and evaluate the Green's function by inserting a complete set of energy eigenfunctions
\begin{align}
  G(\x_f,t_f,\x_i,t_i) = \frac12\int_0^\infty dT \int \frac{d^3 p}{(2\pi)^3}& e^{i\mathbf{p}\cdot(\mathbf{x}_f-\mathbf{x}_i)}\exp\left[-\frac{T}{2}E_p^2\right]\\&\times\sum_\sigma \int d\mathcal{E}\exp\left(T \mathcal{E}\right)\langle t_f|\mathcal{E},\sigma\rangle\langle \mathcal{E},\sigma|t_i\rangle\,,
\end{align}
where $E_p^2 = m^2+p^2$, and the eigenfunctions obey the equation
\begin{equation}
  -\frac12\frac{d^2}{d t^2}\langle t|\mathcal{E},\sigma\rangle - \frac12\omega^2 t^2\langle t|\mathcal{E},\sigma\rangle = \mathcal{E}\langle t|\mathcal{E},\sigma\rangle\,.
\end{equation} 
As before, $\sigma$ labels the two degenerate solutions at energy $\mathcal{E}$, and they should be chosen to be normalized as $\langle \mathcal{E},\sigma|\mathcal{E}',\sigma'\rangle = \delta(\mathcal{E}-\mathcal{E}')\delta_{\sigma\sigma'}$. We will define them so the state with $\sigma=\pm$ describes a wave incoming from $\pm\infty$. The normalization can be found either from the asymptotic form, or by computing the current, which can be obtained from the Wronskian
\begin{equation}
  j=\frac{i}{2}W\left[\langle t|\mathcal{E},\sigma\rangle,\langle t|\mathcal{E},\sigma\rangle^*\right]\,.
\end{equation}
For the appropriately normalized states, the current should be given by $j=-\sigma T/2\pi$ where $T$ is the transmission coefficient $T=1/|\alpha|^2$. Together with our results in Appendix~\ref{app:mass_expansion} for $\alpha$, we see that 
\begin{equation}
  j=-\frac{\sigma}{4\pi}\frac{\exp\left(\frac{\pi\mathcal{E}}{2\omega}\right)}{\cosh(\pi\mathcal{E}/\omega)}\,,
\end{equation} 
which holds if we take the normalized eigenfunctions to be\footnote{To be precise, the phase in the exponential in the argument of the parabolic cylinder function should be taken as $e^{-i({\pi/4}-\delta)}$ for some infinitesimal $\delta>0$.}
\begin{equation}
  \langle t|\mathcal{E},\sigma\rangle = \frac{1}{2^{3/4}\omega^{1/4}\pi}e^{-\frac{\pi\mathcal{E}}{4\omega}}\Gamma\left(\frac12-i\frac{\mathcal{E}}{\omega}\right)D_{-\frac12+i \frac{\mathcal{E}}{\omega}}\left(-\sigma e^{-i\pi/4}\sqrt{2\omega}t\right)\,.
\end{equation}
The Green's function can then be written as
\begin{align}
  G(\x_f,t_f,\x_i,t_i) = \frac12&\int_0^\infty dT \int \frac{d^3 p}{(2\pi)^3} e^{i\mathbf{p}\cdot(\mathbf{x}_f-\mathbf{x}_i)}\exp\left[-\frac{T}{2}E_p^2\right]\nonumber\\&\times \int \frac{d\mathcal{E}}{2\pi}\frac{\exp\left(\frac{\pi\mathcal{E}}{2\omega}\right)}{\sqrt{2\omega}\cosh(\pi\mathcal{E}/\omega)}\exp\left(T \mathcal{E}\right)\nonumber\\
  &\qquad\qquad\times D_{-\frac12+i \frac{\mathcal{E}}{\omega}}\left(-e^{-i\pi/4}\sqrt{2\omega}t_f\right)D_{-\frac12-i \frac{\mathcal{E}}{\omega}}\left(-e^{i\pi/4}\sqrt{2\omega}t_i\right)\nonumber\\
  &\times \int \frac{d\mathcal{E}}{2\pi} \frac{\exp\left(\frac{\pi\mathcal{E}}{2\omega}\right)}{\sqrt{2\omega}\cosh(\pi\mathcal{E}/\omega)}\exp\left(T \mathcal{E}\right)\nonumber\\
  &\qquad\qquad\times D_{-\frac12+i \frac{\mathcal{E}}{\omega}}\left(e^{-i\pi/4}\sqrt{2\omega}t_f\right)D_{-\frac12-i \frac{\mathcal{E}}{\omega}}\left(e^{i\pi/4}\sqrt{2\omega}t_i\right)\,.\label{eq:G_ham_int}
\end{align}
The poles of the integrand are located at $\mathcal{E}=i\omega (n+\frac12)$, and recalling that $T$ has a small positive part, and that the parabolic cylinder functions are entire functions of the index, we see that we can deform the contour into the upper half plane so that the integral can be evaluated by summing over the residues of the poles with $n\geq0$, and the result is 
\begin{align}
  G(\x_f,t_f,\x_i,t_i) = \frac12&\int_0^\infty dT \int \frac{d^3 p}{(2\pi)^3} e^{i\mathbf{p}\cdot(\mathbf{x}_f-\mathbf{x}_i)}\exp\left[-\frac{T}{2}E_p^2\right]\nonumber\\
  &\times \sum_{n=0}^\infty \sqrt{\frac{\omega}{2}}\frac{(-1)^n}{\pi}e^{i\pi/4}e^{i\pi n/2}e^{i(n+1/2)\omega T}\nonumber\\
  &\qquad\qquad\times D_{-n-1}\left(-e^{-i\pi/4}\sqrt{2\omega}t_f\right)D_{n}\left(-e^{i\pi/4}\sqrt{2\omega}t_i\right)\nonumber\\
  &\times \sum_{n=0}^\infty \sqrt{\frac{\omega}{2}}\frac{(-1)^n}{\pi}e^{i\pi/4}e^{i\pi n/2}e^{i(n+1/2)\omega T}\nonumber\\
  &\qquad\qquad\times D_{-n-1}\left(e^{-i\pi/4}\sqrt{2\omega}t_f\right)D_{n}\left(e^{i\pi/4}\sqrt{2\omega}t_i\right)\nonumber\,.
\end{align}
For non-negative integer index, the parabolic cylinder functions are related to the Hermite polynomials 
\begin{equation}
  D_n(\sqrt{2}z) = \frac{1}{2^{n/2}}e^{-z^2/2}H_n(z)\,,
\end{equation}
and consequently obey $D_n(-z)=(-1)^n D_n(z)$. In combination with the identity 
\begin{equation}
  e^{i\pi n/2}D_{-n-1}(iz)+e^{-i\pi n/2}D_{-n-1}(iz)= \frac{\sqrt{2\pi}}{\Gamma(n+1)} D_{n}(z) \,,\label{eq:Dnu_linear}
\end{equation}
we can write the Green's function as
\begin{align}\label{eq:G_ham}
G(\x_f,t_f,\x_i,t_i) = &\frac12\int_0^\infty dT \int \frac{d^3 p}{(2\pi)^3} e^{i\mathbf{p}\cdot(\mathbf{x}_f-\mathbf{x}_i)}\exp\left[-\frac{T}{2}E_p^2-\frac{i}{2}\omega(t_f^2+t_i^2)\right]\\
&\times \sqrt{\frac{\omega}{\pi}}e^{i\pi/4}\sum_{n=0}^\infty e^{i(n+1/2)\omega T} \frac{1}{2^n n!} H_{n}\left(e^{i\pi/4}\sqrt{\omega}t_f\right)H_{n}\left(e^{i\pi/4}\sqrt{\omega}t_i\right)\,.\nonumber
\end{align}
Here we recognize the eigenvalues and eigenfunctions of the harmonic oscillator our potential is related to by analytic continuation. We will return to this form soon and simplify it further to make contact with the form of the Green's function we would obtain in quantum field theory, but before we do, let us evaluate the Green's function using the path integral representation.

\subsection{Path integral representation}
Let us now consider the path integral representation of the Green's function
\begin{multline}
  G(x_f,x_i) = \int\frac{d^3 p}{(2\pi)^3} e^{i\mathbf{p}\cdot(\mathbf{x}_f-\mathbf{x}_i)}\\\times\frac12\int_0^\infty dT \int_{t_i}^{t_f} \mathcal{D} t \exp\left[-\int_0^1 d\tau \, \left(-\frac{1}{2T}\dot{t}^2 + \frac{T}{2}E_p(t)^2\right) \right]\,,
\end{multline}
where the path integral is over all paths $t(\tau)$ with $t(0)=t_i$ and $t(1)=t_f$. The classical path obeys the equation of motion
\begin{equation}
  \ddot{t} + T^2\omega^2 t=0\,,
\end{equation}
and for generic $T$, the solution with appropriate boundary conditions is given by 
\begin{equation}
  t(\tau) = \frac{t_f-t_i\cos(\omega T)}{\sin(\omega T)}\sin(\omega T\tau)+t_i\cos(\omega T\tau)\,.
\end{equation}
Expanding the action around this classical path, we arrive at
\begin{align}
  G(x_f,x_i) = \int&\frac{d^3 p}{(2\pi)^3} e^{i\mathbf{p}\cdot(\mathbf{x}_f-\mathbf{x}_i)}\nonumber\\&\times\frac12\int_0^\infty dT \exp\left[-\frac{T}{2}E_p^2+\frac12\omega (t_f^2+t_i^2)\cot(\omega T)-\omega \frac{t_i t_f}{\sin(\omega T)}\right]\nonumber\\
   &\qquad\qquad\qquad\quad \times\int \mathcal{D} \delta t \exp\left[-\int_0^1 d\tau \, \left(-\frac{1}{2T}\delta\dot{t}^2 + \frac{T}{2}\omega^2\delta t^2\right) \right]\,,
\end{align}
where as before $E_p^2=m^2+p^2$ and the paths now obey Dirichlet boundary conditions $\delta t(0)=\delta t(1)=0$, and the measure is defined so that 
\begin{equation}
  \int \mathcal{D} \delta t \exp\left[-\int_0^1 d\tau \, \frac{1}{2T}\delta\dot{t}^2 \right]=\frac{1}{2\pi T}\,.
\end{equation}
The resulting functional determinant can be evaluated using $\zeta$-function regularization and using the Weierstrass product formula for the sine function, or by using the Gelfand-Yaglom method. The resulting expression for the Green's function is
\begin{align}
  G(x_f,x_i) = \int&\frac{d^3 p}{(2\pi)^3} e^{i\mathbf{p}\cdot(\mathbf{x}_f-\mathbf{x}_i)}\nonumber\\&\times\frac{i}{2}\sqrt{\frac{\omega}{2\pi}}\int_0^\infty dT \frac{\exp\left[-\frac{T}{2}E_p^2+\frac12\omega (t_f^2+t_i^2)\cot(\omega T)- \frac{\omega}{\sin(\omega T)}t_i t_f\right]}{\sqrt{\sin(\omega T)}}\,.
\end{align}
In order to make contact with our result in the previous subsection, let us define $q=\exp(i\omega T)$ and write the Green's function as 
\begin{align}
  G(x_f,x_i) = \frac{1}{2}\int_0^\infty dT\int&\frac{d^3 p}{(2\pi)^3} e^{i\mathbf{p}\cdot(\mathbf{x}_f-\mathbf{x}_i)}\exp\left[-\frac{T}{2}E_p^2-\frac{i}{2}\omega (t_f^2+t_i^2)\right]\nonumber\\&\times \sqrt{\frac{\omega}{\pi}} e^{i\pi/4}\sqrt{q}\frac{\exp\left[-i\omega (t_f^2+t_i^2)\frac{q^2}{1-q^2} + 2i\omega t_i t_f\frac{q}{1-q^2}\right]}{\sqrt{1-q^2}}\,.
\end{align}
Comparing this result to our result derived in the Hamiltonian formulation by inserting a complete set of energy eigenfunctions~\eqref{eq:G_ham}, we see that we must have 
\begin{align}
  \frac{1}{\sqrt{1-q^2}}&\exp\left[-i\omega (t_f^2+t_i^2)\frac{q^2}{1-q^2} +  2i\omega t_i t_f\frac{q}{1-q^2}\right]\nonumber \\ &\qquad\qquad\qquad = \sum_{n=0}^\infty \frac{q^n}{2^n n!} H_{n}\left(e^{i\pi/4}\sqrt{\omega}t_f\right)H_{n}\left(e^{i\pi/4}\sqrt{\omega}t_i\right) \,.
\end{align}
This relation is known as Mehler's formula. One can also make contact between the two formulations by recognizing the expression for the bilinear generating functional for parabolic cylinder functions~\cite{Erdelyi} in equation~\eqref{eq:G_ham_int}.

\subsection{Relation to quantum field theory form of propagator}
\begin{figure}[t!]
\centering
\includegraphics{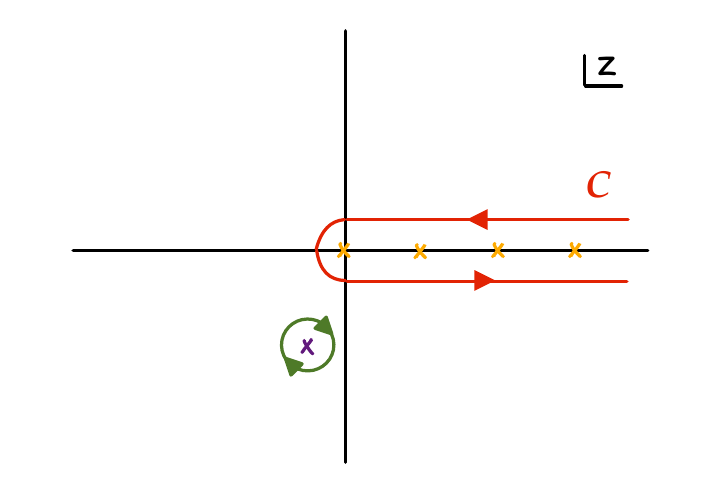}
\caption{The contour $\mathcal{C}$ in the complex $z$-plane used in the evaluation of the Green's function is shown in red. The orange crosses indicated the positions of the poles of the $\Gamma$-function, and the purple cross indicates the pole originating from the integral over $T$. Since the parabolic cylinder functions are entire functions of their arguments, and the $\Gamma$-function decays exponentially in the upper and lower half-plane, the contour can be deformed to the green contour.}
\label{fig:contour}
\end{figure}

So far our expressions still look different from the form of the two-point correlation function we encountered in section~\ref{sec:propagator}. To make contact, let us take equation~\eqref{eq:G_ham} as our starting point. In terms of parabolic cylinder functions, we can write the Green's function as
\begin{align}
  G(\x_f,t_f,\x_i,t_i) = &\frac12\int_0^\infty dT \int \frac{d^3 p}{(2\pi)^3} e^{i\mathbf{p}\cdot(\mathbf{x}_f-\mathbf{x}_i)}\exp\left[-\frac{T}{2}E_p^2\right]\\
  &\times \sqrt{\frac{\omega}{\pi}}e^{i\pi/4}\sum_{n=0}^\infty e^{i(n+1/2)\omega T} \frac{(-1)^n}{\Gamma(n+1)} D_{n}\left(e^{i\pi/4}\sqrt{2\omega}t_f\right)D_{n}\left(-e^{i\pi/4}\sqrt{2\omega}t_i\right)\,.\nonumber
  \end{align} 
We will now write this as a contour integral, and we have used the identity $D_n(z)=(-1)^nD_n(-z)$, before doing so to ensure we are free to deform the contour as desired. 
\begin{align}
  G(\x_f,t_f,\x_i,t_i) = &\frac12\int_0^\infty dT \int \frac{d^3 p}{(2\pi)^3} e^{i\mathbf{p}\cdot(\mathbf{x}_f-\mathbf{x}_i)}\exp\left[-\frac{T}{2}E_p^2+\frac{i}{2}\omega T\right]\\
  &\times \sqrt{\frac{\omega}{\pi}}e^{i\pi/4}\int_\mathcal{C} \frac{dz}{2\pi i}\frac{\pi}{\sin(\pi z)} \frac{1}{\Gamma(z+1)} e^{i z \omega T} D_{z}\left(e^{i\pi/4}\sqrt{2\omega}t_f\right)D_{z}\left(-e^{i\pi/4}\sqrt{2\omega}t_i\right)\,,\nonumber
\end{align}
where the contour $\mathcal{C}$ is shown in Figure~\ref{fig:contour}. 
Using the identity $\Gamma(1+z)\Gamma(-z)=-\pi/\sin(\pi z)$, we can write the Green's function as
\begin{align}
  G(\x_f,t_f,\x_i,t_i) = &\frac12\int_0^\infty dT \int \frac{d^3 p}{(2\pi)^3} e^{i\mathbf{p}\cdot(\mathbf{x}_f-\mathbf{x}_i)}\exp\left[-\frac{T}{2}E_p^2+\frac{i}{2}\omega T\right]\\
  &\times - \sqrt{\frac{\omega}{\pi}}e^{i\pi/4}\int_\mathcal{C} \frac{dz}{2\pi i}\Gamma(-z) e^{i z \omega T} D_{z}\left(e^{i\pi/4}\sqrt{2\omega}t_f\right)D_{z}\left(-e^{i\pi/4}\sqrt{2\omega}t_i\right)\,,\nonumber
\end{align}
Since the parabolic cylinder functions are entire functions of $z$, the only poles at this point are those arising from the $\Gamma$-function located at non-negative integers. We can now perform the integral over $T$. This leads to an additional pole in the complex $z$ plane at $z=-\bar\nu-1=-1/2-i E_p^2/2\omega$, and since the parabolic cylinder functions are entire functions of the index and the $\Gamma$-function decays exponentially in the upper and lower half plane, \mbox{$|\Gamma(x+i y)|\sim\sqrt{2\pi}|y|^{x-1/2}e^{-\pi|y|}$}, we can deform the contour so that the integral is given by the residue at this pole 
\begin{align}
  G(\x_f,t_f,\x_i,t_i) = &\int \frac{d^3 p}{(2\pi)^3} e^{i\mathbf{p}\cdot(\mathbf{x}_f-\mathbf{x}_i)}\\
  &\times e^{-i\pi/4}\frac{\Gamma(\bar\nu+1)}{\sqrt{2\pi}\sqrt{2\omega}}  D_{-\bar\nu-1}\left(e^{i\pi/4}\sqrt{2\omega}t_f\right)D_{-\bar\nu-1}\left(-e^{i\pi/4}\sqrt{2\omega}t_i\right)\,.\nonumber
\end{align}
For $t_f$ and $t_i$ long after the production event, the phase of the argument of the second parabolic cylinder function does not satisfy $|\arg(z)|<3\pi/4$, and we can use the identity~\eqref{eq:Dnu_linear}, this time in the form 
\begin{equation}
  D_{-\bar\nu-1}(z) = \frac{\sqrt{2\pi}}{\Gamma(\bar\nu+1)}e^{i\pi\bar{\nu}/2} D_{\bar\nu}(i z)- e^{i\pi\bar\nu}D_{-\bar\nu-1}(-z)\,,\label{eq:Dnu_rel}
\end{equation}
to write the Green's function as 
\begin{align}
  G(\x_f,t_f,\x_i,t_i) = &\int \frac{d^3 p}{(2\pi)^3} e^{i\mathbf{p}\cdot(\mathbf{x}_f-\mathbf{x}_i)}\\
  &\times \left[e^{-i\pi/4} e^{i\pi\bar{\nu}/2}\frac{1}{\sqrt{2\omega}}  D_{-\bar\nu-1}\left(e^{i\pi/4}\sqrt{2\omega}t_f\right) D_{\bar\nu}\left(e^{-i\pi/4}\sqrt{2\omega}t_i\right)\right.\nonumber\\
  &\quad-\left. e^{-i\pi/4}e^{i\pi\bar\nu} \frac{\Gamma(\bar\nu+1)}{\sqrt{2\pi}\sqrt{2\omega}} D_{-\bar\nu-1}\left(e^{i\pi/4}\sqrt{2\omega}t_f\right)D_{-\bar\nu-1}(e^{i\pi/4}\sqrt{2\omega}t_i)\right] \,.\nonumber
\end{align}
Using the results from Appendix~\ref{app:exact}, we recognize this as
\begin{align}
  G(\x_f,t_f,\x_i,t_i) = &-i\int \frac{d^3 p}{(2\pi)^3} e^{i\mathbf{p}\cdot(\mathbf{x}_f-\mathbf{x}_i)}\left[u_{\rm out}(t_f)u^*_{\rm out}(t_i)+\frac{\beta_p}{\alpha^*_p}u_{\rm out}(t_f)u_{\rm out}(t_i)\right]\nonumber\\=&-i\langle\chi(t_f)\chi(t_i)\rangle\,.
\end{align}
So we have obtained the Green's function in a first quantized Hamiltonian and path integral formulation, and we have seen how these results are related to the expression most easily obtained in quantum field theory.

\section{Ratio of functional determinants in WKB approximation}\label{app:ratio}
In this appendix, we use the WKB approximation to evaluate the ratio of functional determinants 
 \begin{equation}
  \sqrt{\frac{\det' M_0}{\det' M}}\,,
\end{equation}
we encountered in section~\ref{ssec:particle prefactor} with
\begin{equation}
  M_0(\tau,\tau') = \left(-\frac{1}{T_\star^2}\frac{\partial^2}{\partial \tau^2} \right)\delta(\tau-\tau')\,,
\end{equation} and 
\begin{equation}
  M(\tau,\tau') = \left(-\frac{1}{T_\star^2}\frac{\partial^2}{\partial \tau^2} - \omega^2(\tau)\right)\delta(\tau-\tau')+\frac{4}{S_\star T_\star^3}\ddot{\bar{x}}^0(\tau)\ddot{\bar{x}}^0(\tau')\,,
\end{equation}
acting on functions with periodic boundary conditions. Let us first consider the spectrum and eigenfunctions of 
\begin{equation}
  \hat{M}(\tau,\tau') = \left(-\frac{1}{T_\star^2}\frac{\partial^2}{\partial \tau^2} - \omega^2(\tau)\right)\delta(\tau-\tau')\,.
\end{equation}
In the WKB approximation, the normalized mode functions are 
\begin{eqnarray}
  \delta x^{(1)}_k(\tau)&=&\sqrt{\frac{2}{T_\star}}\sqrt{\frac{\bar{\omega}}{\omega(\tau)}}\sin\left(T_\star\int_0^\tau d\tau'\sqrt{\omega^2(\tau')+\lambda_k}\right)\,,\\
  \delta x^{(2)}_k(\tau)&=&\sqrt{\frac{2}{T_\star}}\sqrt{\frac{\bar{\omega}}{\omega(\tau)}}\cos\left(T_\star\int_0^\tau d\tau'\sqrt{\omega^2(\tau')+\lambda_k}\right)\,,
\end{eqnarray}
where we have defined
\begin{equation}
  \bar{\omega}^2=\int_0^1 d\tau \,\omega^2(\tau)\,,
\end{equation}
and have expanded $\omega^2(\tau)=\bar{\omega}^2+\delta\omega^2(\tau)$. Using that 
\begin{equation}
  \int_0^1 d\tau \delta\omega^2(\tau)=0\,,
\end{equation}
we see that imposing periodicity implies that, up to terms of second order in $\delta\omega^2$, the eigenvalues $\lambda_k$ are given by
\begin{equation}
  \lambda_k = \left(\frac{2\pi }{T_\star}\right)^2\left(k^2-n^2\right)\,.
\end{equation}
Here we have used our knowledge that the mode with $k=n$ is a zero mode to write  $\bar{\omega} =2\pi n/T_\star$. As expected, the eigenvalues are negative for $k<n$, positive for $k>n$, and $\delta x^{(1)}_n(\tau)$ and $\delta x^{(2)}_n(\tau)$ with $k=n$ are zero modes of $\hat{M}(\tau,\tau')$.   

Let us now consider the effect of the rank one operator in $M(\tau,\tau')$ on the eigenvalues. Taking a derivative of equation~\eqref{eq:zeromode}, we see that
\begin{equation}
  \ddot{\bar{x}}^0(\tau)= -i \sqrt{S_\star} T_\star^2\omega(\tau)\,\delta x^{(2)}_n(\tau)\,, 
\end{equation}
so that $M(\tau,\tau')$ becomes
\begin{equation}
  M(\tau,\tau') = \left(-\frac{1}{T_\star^2}\frac{\partial^2}{\partial \tau^2} - \omega^2(\tau)\right)\delta(\tau-\tau')-4 T_\star \omega(\tau) \omega(\tau') \delta x^{(2)}_n(\tau)\delta x^{(2)}_n(\tau')\,.
\end{equation}

Up to terms of second order in $\delta\omega^2(\tau)$we see that in the WKB approximation, the eigenvalues of $M(\tau,\tau')$ are the same as those of $\hat{M}(\tau,\tau')$ for all modes except $\delta x^{(2)}_n(\tau)$. For this mode, the eigenvalue is 
\begin{equation}
 \lambda_n^{(2)}= \int_0^1\int_0^1 d\tau d\tau' T_\star\delta x^{(2)}_n(\tau) M(\tau,\tau')\delta x^{(2)}_n(\tau') = -4 \bar{\omega}^2\,.
\end{equation}
Recalling $\bar{\omega}T_\star=2\pi n$, we see that the spectrum of $M(\tau,\tau')$ is given by 
\begin{eqnarray}
  \lambda^{(1)}_k &=& \left(\frac{2\pi }{T_\star}\right)^2\left(k^2-n^2\right)\,,\qquad k=1,\dots,\infty\,,\\
  \lambda^{(2)}_k &=& \left(\frac{2\pi }{T_\star}\right)^2\left(k^2-n^2\right)\,,\qquad k=0,\dots,\infty\quad k\neq n\,,\\
  \lambda^{(2)}_n &=& -4 \left(\frac{2\pi n}{T_\star}\right)^2\,.
\end{eqnarray}
Given the spectrum, we see that the ratio of functional determinants can be written as
\begin{equation}
  \sqrt{\frac{\det' M_0}{\det' M}} = - \frac12\left[\prod_{\substack{k=1\\ k\neq n}}^\infty 1-\left(\frac{n}{k}\right)^2\right]^{-1}\,.
\end{equation} 
Setting $z=n+\epsilon$ in Euler's formula for the sine 
\begin{equation}
  \sin(\pi z) = \pi z\prod_{k=1}^\infty \left(1-\frac{z^2}{k^2}\right)\,,
\end{equation}
and taking the limit $\epsilon\to 0$, we find that the infinite product is given by 
\begin{equation}
  \prod_{\substack{k=1\\ k\neq n}}^\infty 1-\left(\frac{n}{k}\right)^2 = \frac{(-1)^{n-1}}{2} \,,
\end{equation}
so that the square root of the ratio of functional determinants is 
\begin{equation}
  \sqrt{\frac{\det' M_0}{\det' M}} = -(-1)^{n-1}\,.
\end{equation}

\section{Choice of time slice}\label{app:time}
Here we revisit the question of selecting the appropriate time slice in \labelcref{qft_wf}. The time dependence is contained in the $u^{\rm out}_p(t)$'s, given in \labelcref{uoutmodefnc}, and by taking the Fourier transform numerically, we can obtain $\Psi(r,t)$ as a function of the time slice. We can find the peak separation at each $t$, by finding the $r$ where $|\Psi(r,t)|^2\, dV$ attains a maximum. Physically, we expect the average separation should undergo some nonlinear transient behavior between the maxima of the nonadiabaticity parameter $|\dot{\omega}_p/\omega_p^2|$, when particle production is taking place. For the quadratic time-dependence $\omega_p^2 = E_p^2 +\omega^2 t^2$ where $E_p=\sqrt{m_0^2+p^2}$, this ratio is $|\omega^2 t/(E_p^2 + \omega^2 t^2)^{3/2}|$. After some time $t\,\gsim \,t_{+} = E_p/\sqrt{2}\omega$, the particles become well-defined and start propagating freely. Eventually, the time dependence of their mass will start slowing them down and the average separation will evolve approximately as $\hat{r}\sim \dfrac{pt}{\sqrt{m_0^2+\omega^2t^2}}$. The quantity we wish to evaluate numerically is
\begin{align}
P &\equiv r^2 |\Psi(r,t)|^2 = r^2\left|\int \frac{d^3p}{(2\pi)^3} e^{i \mathbf{p}\cdot \mathbf{r}}\,\frac{\beta_p^*}{\alpha_p^*} u^{\rm out}_p(t)^2\right|^2 = r^2\left|\int_0^\infty \frac{p dp}{2\pi^2} \, \frac{\sin pr}{r}\,\frac{\beta_p^*}{\alpha_p^*} u^{\rm out}_p(t)^2\right|^2 \nonumber \\
&= \frac{e^{-S_{\mathrm{inst}}}}{16\pi^5 \omega}\left|\int_0^\infty dp\, p \sin pr\,e^{-\frac{p^2\pi}{2\omega}}\,\Gamma\left(\frac{1}{2}+\frac{iE_p^2}{2\omega}\right)\,D_{-\frac{1}{2} - \frac{iE_p^2}{2\omega}}\left(e^{i\pi/4}\sqrt{2\omega}\,t\right)^2\right|^2\,.
\end{align}
\begin{figure}[t]
  \centering
  \includegraphics[width=0.8\textwidth]{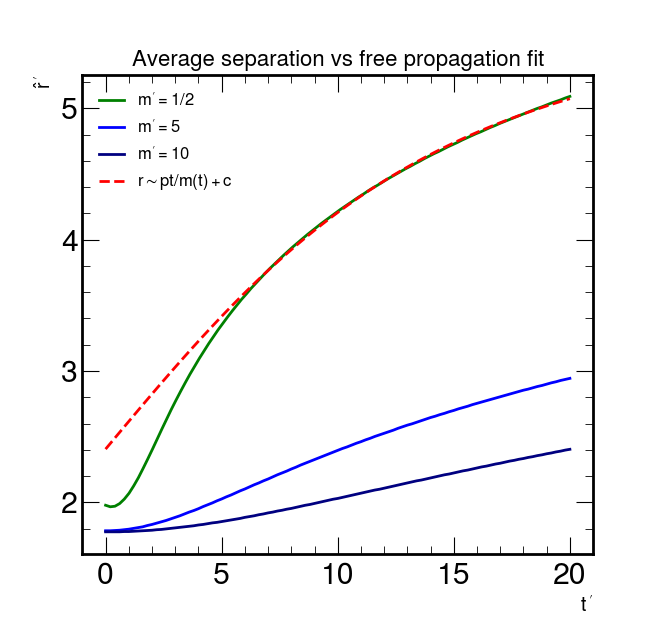}
  \caption{Peak of the distribution $|\Psi(r,t)|^2\, dV$ at different time slices, for different mass ratios $m'= m_0/\sqrt{\omega}$. The $m'=5$ and 10 cases correspond to the large $S_\mathrm{inst}$ limit where the wavefunction is well-approximated by~\eqref{eq:gaussianwavefunction}, and $\hat{r}' \approx 1.78$, matching~\eqref{rhat particle}.}
  \label{fig:time_dep}
\end{figure}
We can make the integrand dimensionless by scaling out the (mass dimension 2) parameter $\omega$ and defining $p' =p/\sqrt{\omega}, \; r' = r\sqrt{\omega}, \; t' = t\sqrt{\omega}$. Parametrically, the integral now depends only on the mass ratio $m' = m_0/\sqrt{\omega}$ (which is nothing but $\sqrt{S_\mathrm{inst}}/ \pi$), through the rescaled $E'_p = \sqrt{m'^2 + p'^2}$:
\begin{gather}
P = \frac{\omega e^{-S_{\mathrm{inst}}}}{16\pi^5}\left|\int_0^\infty dp' p' \sin p'r'\,e^{-\frac{p'^2\pi}{2}}\,\Gamma\left(\frac{1}{2}+\frac{i{E'_p}^2}{2}\right)\,D_{-\frac{1}{2} - \frac{iE'^2_p}{2}}\left(e^{i\pi/4}\sqrt{2}t'\right)^2 \right|^2 \\
\hat{r}'(t) = \underset{r'}{\mathrm{argmax}} \; P
\end{gather}
The corresponding separation $\hat{r}'(t)$ is plotted in Figure \ref{fig:time_dep} for some representative examples of the mass ratio. The separation is minimized close to $t=0$, which justifies our use of the $t=0$ slice as a conservative estimate, and $\Psi(r,0)$ captures all the relevant parametric dependence. As expected, there is some transient behavior and the separation increases non-linearly, until eventually the particles start moving as free particles with fixed $p$ and time dependent mass. This happens more slowly for particles with a larger mass ratio, as the bare mass $m_0$ dominates the kinematics.

\pagebreak

\bibliographystyle{JHEP}
\bibliography{refs}
\end{document}